\pdfoutput=1
\documentclass[preprint]{aastex}
\usepackage{hyperref}
\usepackage{cleveref}

\usepackage[margin=1.0 in]{geometry}
\usepackage{times}
\usepackage{amsmath}
\usepackage{rotating,tabularx}

\newcommand{\bc}{\begin{center}}
\newcommand{\ec}{\end{center}}
\newcommand{\be}{\begin{equation}}
\newcommand{\ee}{\end{equation}}
\newcommand{\bfig}{\begin{figure}}
\newcommand{\efig}{\end{figure}}
\newcommand{\m}{\mbox}
\newcommand{\Ms}{M$_\odot$}

\extrafloats{100}

\received{?}
\revised{?}
\accepted{?}
	
\usepackage{amssymb}
\begin{document}
\title{Atmospheres and UV Environments of Earth-like Planets Throughout Post-Main Sequence Evolution}
\author{Thea Kozakis, Lisa Kaltenegger}
\affil{Carl Sagan Institute, Cornell University, Ithaca, NY 14853}
\date{}

\begin{abstract}
During the post-main sequence phase of stellar evolution the orbital distance of the habitable zone, which allows for liquid surface water on terrestrial planets, moves out past the system's original frost line, providing an opportunity for outer planetary system surface habitability. We use a 1D coupled climate/photochemistry code to study the impact of the stellar environment on the planetary atmospheres of Earth-like planets/moons throughout its time in the post-main sequence habitable zone. We also explore the ground UV environments of such planets/moons and compare them to Earth's. We model the evolution of star-planet systems with host stars ranging from 1.0 to 3.5 \Ms\ throughout the post-main sequence, calculating stellar mass loss and its effects on planetary orbital evolution and atmospheric erosion. The maximum amount of time a rocky planet can spend continuously in the evolving post-MS habitable zone ranges between~56 and 257~Myr for our grid stars. Thus, during the post-MS evolution of their host star, subsurface life on cold planets and moons could become remotely detectable once the initially frozen surface melts. Frozen planets or moons, like Europa in our Solar System, experience a relatively stable environment on the horizontal branch of their host stars' evolution for millions of years. 
\end{abstract}

\maketitle

\section{Introduction}

As a star evolves, the orbital distance where liquid water is possible on the surface of an Earth-like planet, the so-called habitable zone (HZ), evolves as well. While stellar properties are relatively stable on the main sequence (MS), post-MS evolution of a star involves significant changes in stellar temperature and radius, which is reflected in the changing irradiation at a specific orbital distance during the red giant branch (RGB), and for stars massive enough ($>$~$\sim$0.5~\Ms), the horizontal branch (HB) and the asymptotic giant branch (AGB). 

To explore post-MS planetary systems for signs of life, it is critical to understand how the host star influences signs of life in the atmosphere during post-MS evolution. Not only do the host star's drastically changing temperature and luminosity change the orbital distance of the HZ, but stellar mass loss can also erode a planet's atmosphere through resulting stellar winds. The majority of studies on the habitability of rocky planets and moons have focused on planetary systems around MS stars, with only some work done on habitability around pre- and post-MS stars (see, e.g.\ \cite{ster03,lope05,danc13,rami14,rami16}), or stellar remnants (e.g.\ \cite{wols92,barn13,koza18}). However, no study has explored the evolution of the atmospheres and potential detectable biosignatures or UV surface environment for such post-MS planets yet.

Liquid surface water is used because it remains to be demonstrated whether subsurface biospheres, for example, under an ice layer on a frozen planet, can modify a planet's atmosphere in ways that can be detected remotely. Although planets located outside the HZ are not excluded from hosting life, detecting biosignatures remotely on such planets should be extremely difficult (see also \cite{rami18r}). During post-MS evolution, the orbital distance of the HZ often moves out past the system's original frost line, where water can remain in frozen form solely based on stellar irradiation at this orbital distance. In our solar system 99.99\% of H$_2$O is located beyond the frost line \citep{ster03}, presenting the opportunity for surface habitability of initially frozen planets and moons in the outer solar system during the star's post-MS evolution.

As of now, there are tens of known gaseous planets around red giant (RG) stars (e.g.\ \cite{jone14}), although a terrestrial planet orbiting an RG has yet to be discovered. However, due to the wide separation of the post-MS HZ, such planets could be resolved via direct imaging by upcoming telescopes like the Extremely Large Telescope (ELT; see also \cite{rami16}).

For planets orbiting post-MS stars, several studies suggest the possibility of life developing on planets orbiting subgiant or RG hosts (e.g.\ \cite{lope05}), which strongly depends on the time required for life to evolve compared to the star's post-MS lifetime. \cite{rami16} have added the possibility that ice on initially frozen planets and moons could melt during the RG phase of their host stars, revealing life that initially developed under the surface.

In this study, we explore the influence of post-MS stellar irradiation on the climate, as well as its surface UV environment for Earth-like planets in the post-MS HZ. We focus on host stars with masses of~1.0 to 3.5~\Ms, which can undergo the post-MS phase in a galaxy the age of our Milky Way. We use observed and modeled spectra of close-by RGs for our stellar models. Some of these RGs can be placed onto one evolutionary track, thus giving us first insight into a planet's changing environment during the post-MS phase of its host star. 

We first model the joint evolution of the star and the orbital distance of its post-MS HZ, along with stellar mass loss and its impacts on planetary orbital radii and atmospheric erosion following \cite{rami16}, although with an extension to higher-mass stars (1.9 to 3.5~\Ms).  We then model the atmospheres of Earth-like planets in the post-MS HZ using a 1D coupled climate/photochemistry code to explore the chemistry and UV surface environment of our model planets in the post-MS HZ. We focus on the change of atmospheric signatures that can indicate life on a planet: ozone and oxygen in combination with a reducing gas like methane or N$_2$O (see, e.g.\ the review by \cite{kalt17} for details). We also show climate indicators for our model planets like water and CO$_2$, which, in addition, can indicate whether the oxygen production can be explained abiotically. Section~\ref{methods} explains our methods, Section~\ref{results} shows our results, and Sections~\ref{discussion} and~\ref{conclusions} discuss and conclude our work.

\section{Methods \label{methods}}

We model the evolution of post-MS host stars and their effects on planetary orbits for 1 Earth-mass planets to calculate how long such a planet could stay within the boundaries of the post-MS HZ, as well as the effect of the host star's evolution on the atmospheres and UV surface environment of our model planets during the time in the post-MS HZ. 

\begin{table}[h!]
\centering
\caption{Padova catalog stellar lifetimes\label{lifetimes}}
\begin{tabular}{crrrrrrr}
\hline
Mass & MS & Post-MS &RGB & HB & AGB \\
(\Ms)		 & (Myr)	 & (Myr) & (Myr)  &(Myr)  & (Myr)  \\
 \hline
1.0	& 11680	& 1011	& 851	& 133	& 27	\\
1.3	& 4348	& 663	& 529	& 124	& 10	\\
1.5	& 2901	& 294	& 157	& 125	& 12	\\
1.7	& 1964	& 236	& 88 		& 136	& 12	\\
1.9	& 1409	& 238	& 52		& 171	& 15	\\
2.0	& 1205	& 260	& 38  	& 206	& 16	\\
2.3	& 822	& 260	& 9		& 226	& 25	\\
2.5	& 651	& 201	& 6		& 175	& 20	\\
3.0	& 396	& 104	& 2		& 92		& 10	\\
3.5	& 263	& 58		& 1		& 51		& 6  \\
 \hline
 \end{tabular}
 \end{table}

\subsection{Post-MS HZ boundaries}

We use stellar evolutionary tracks from the Padova catalog \citep{bert08,bert09} that model stellar evolution from the zero-age MS (ZAMS) to the first significant thermal pulse on the AGB. All tracks used in this study have a solar-like metallicity with Z~=~0.017 and Y~=~0.26, from which we obtain the changing luminosity, temperature, and surface gravity, as well as the predicted time points for the beginnings of the RGB, HB, and AGB.

As shown in Table~\ref{lifetimes}, the timescales for the phases of stellar evolution are highly dependent on stellar mass. While MS lifetimes are always longer for smaller masses, post-MS timescales, particularly the length of the relatively stable HB, do not linearly correlate with host mass. Stars below 2~\Ms\ (0.8 to 2.0~\Ms) undergo a helium flash, which greatly increases the luminosity in the post-MS phase. 
Stars that do not undergo the helium flash experience a less extreme luminosity change during the RGB phase and spend a significantly higher percentage of their post-MS lifetimes on the HB.  As a result, the absolute length of the HB peaks for stellar masses of 2.3~\Ms\ (see Table~\ref{lifetimes} and Figure~\ref{HR_fitting}).

The HZ is defined as the region around one or multiple stars in which liquid water could be stable on an Earth-like rocky planet's surface (e.g.\ \cite{kast93,kalt13,kane13}), facilitating the remote detection of possible atmospheric biosignatures. The width and orbital distance of a given HZ depend to a first approximation on two main parameters: incident stellar flux and planetary atmospheric composition. The incident stellar flux depends on the stellar luminosity and stellar spectral energy distribution, the planet's orbital distance (semimajor axis), and the eccentricity of the planetary orbit. The warming due to atmospheric composition depends on the planet's atmospheric makeup, energy distribution, and resulting albedo and greenhouse warming.

In the literature, very different values of stellar irradiance are used as boundaries for the HZ (see review \cite{kalt17}). Here we use the empirical HZ boundaries, based on the solar flux received by our neighboring rocky planets, Venus and Mars, when we can exclude liquid water on their surfaces. This recent Venus and early Mars irradiation and the resulting HZ limits were originally defined using a 1D climate model by \cite{kast93} and updated in \cite{kopp13} and \cite{rami17,rami18} for MS stars with effective temperatures ($T_{\footnotesize \mbox{eff}}$) between~2,600 and 10,000~K. Note that the inner limit of the empirical HZ is not well known because of the lack of a reliable geological surface history of Venus beyond about 1~billion years due to resurfacing of the stagnant lid, which allows for the possibility of a liquid surface ocean.  However, it does not stipulate a liquid ocean surface.

We calculate the flux boundaries ($S_{\footnotesize \mbox{eff}}$) of the post-MS HZ boundaries for the evolving stellar luminosity (Table~\ref{lifetimes}) during the post-MS.  Equation~\ref{HZ_eq} gives a third-order polynomial curve fit of the modeling results for host stars as shown in \cite{kalt17}, based on values derived from models by \cite{kast93} and \cite{kopp13,kopp14}, and an extension of that work to 10,000~K by \cite{rami18}.  The flux values of the HZ are defined by
\begin{equation}
S_{\footnotesize \mbox{eff}} = S_{\footnotesize \mbox{Sun}} + aT* + bT*^{2} + cT*^3 + dT*^4,
\label{HZ_eq}
\end{equation}
where $T* = T_{\footnotesize \mbox{eff}} - 5780$, and $S_{\footnotesize \mbox{Sun}}$ is the stellar incident values at the HZ boundaries in our solar system. Table~\ref{HZ_co} \citep{rami18} shows the constants $a$, $b$, $c$, and $S_{ \mbox{\footnotesize Sun}}$ needed to derive the stellar flux at the HZ limits valid for $T_{\footnotesize \mbox{eff}}$ between 2,600 to 10,000~K. The inner boundaries of the empirical HZ (recent Venus), as well as an alternative inner edge limit for 3D Global Climate models (3D; \cite{leco13}) and the outer limits (early Mars), are all included. The outer HZ limit in 3D and 1D models are consistent and therefore not given in separate columns in Table~\ref{HZ_co} (see e.g.\ \cite{turb17,wolf18}). However, climate models still show limitations due to unknown cloud feedback for higher stellar irradiation. Thus, we use the empirical HZ based on recent Venus and early Mars flux limits for our calculations here, even though the inner limit is uncertain, as explained before. Table~\ref{HZ_co} provides values to estimate the size of the empirical HZ, as well as models of the HZ based on 3D atmospheric models for a planet orbiting our Sun \citep{leco13} and adapted for different host stars \citep{rami14}.

The orbital distance of the HZ boundaries around a star with luminosity $L$ can be calculated from the incident stellar flux $S_{\footnotesize \mbox{eff}}$ using
\begin{equation}
d = \sqrt \frac{L/L_\odot  }{S_{\footnotesize \mbox{eff}}},
\label{l_eq}
\end{equation}
with $L$ measured in solar units ($L_\odot$) and the orbital distance $d$ in astronomical units.

\begin{table}[h!]
\begin{center}
\caption{HZ constants from \cite{rami18} \label{HZ_co}}
\begin{tabular}{lrrr}
\hline
Constants & Recent Venus limit & 3D model limit & Early Mars \\
		& (inner edge)	& (inner edge) & (outer edge)  \\
\hline
$S_{\footnotesize \mbox{Sun}}$	& 1.7665		& 1.1066		& 0.324		\\
$a$		& 1.335E-4	& 1.2181E-4	& 5.3221E-5	\\
$b$		& 3.1515E-9	& 1.534E-8	& 1.4288E-9	\\
$c$		& -3.3488E-12	& -1.5018e-12	& -1.1049E-12	\\
\hline
\end{tabular}
\end{center}
\end{table}

\subsection{Planetary semimajor axis evolution}

Significant stellar mass loss occurs on the RGB and AGB, both impacting the orbital radii of orbiting planets, as well as potentially eroding planetary atmospheres.  We model stellar mass loss on the RGB ($\dot{M}_{RGB}$ in $M_\odot$ yr$^{-1}$) using the modified Reimers equation \citep{reim75,vass93} and the \cite{baud83} parameterization for the AGB ($\dot{M}_{AGB}$ in $M_\odot$ yr$^{-1}$) following  \cite{rami16},

\begin{equation}
\dot{M}_{RGB} = -\frac{4}{3} \times 10^{-13} \frac{L}{gR_{\footnotesize \mbox{star}}},
\label{RG_m}
\end{equation}
\begin{equation}
\dot{M}_{AGB} = -\frac{4} \times 10^{-13} \frac{M_i}{M_{star}}\frac{L}{gR_{\footnotesize \mbox{star}}},
\label{AG_m}
\end{equation}
where $L$ is the stellar luminosity, $g$ is gravity, and $R$ is the radius, and $M_i$ and $M_{\footnotesize \mbox{star}}$ are the initial and current stellar mass, with everything in Solar units.

Orbital variation that will occur as a result of the host star's mass loss can  be approximated as (assuming the planetary mass is negligible compared to its host's mass)
\begin{equation}
\frac{1}{D}\frac{ dD}{dt} = -\frac{1}{M_{\footnotesize \mbox{star}}}\frac{dM_{\footnotesize \mbox{star}}}{dt},
\end{equation}
where $D$ is the orbital distance.  This can be integrated to obtain
\begin{equation}
D(t) = D_o \frac{M_{i}}{M_{\footnotesize \mbox{star}}(t)},
\label{D_t}
\end{equation}
where $D_o$ is the initial orbital distance and $M_i$ is the initial stellar mass.  Using Equations~\ref{RG_m} and~\ref{AG_m} to model $M_{\footnotesize \mbox{star}}$($t$), we calculate the planet's $D$($t$) for a given $D_o$.

\subsection{Planetary atmospheric erosion}

We model planetary atmospheric erosion following \cite{rami16} using the formalism from \cite{cant91} to model planetary atmospheric erosion from stellar winds caused by mass loss. Planetary atmospheric loss per unit time ($\dot{M}_a$) from the stellar winds is approximated by
\begin{equation}
\dot{M}_a \approx - 2 \alpha \pi R^2_p \rho_w v_w .
\end{equation}
In the stellar wind flow, $v_w$ is sound speed, $\rho_w$ is the density, and $\alpha$ is the entrainment efficiency.  Using the relation of 
\begin{equation}
\dot{M}_{\footnotesize \mbox{star}} = 4 \pi D^2 \rho_w v_w,
\end{equation}
we can simplify $\dot{M}_a$ as
\begin{align}
\dot{M}_a \approx - 2 \alpha \pi R^2_p \frac{\dot{M}_{\footnotesize \mbox{star}}}{4\pi D^2} \\
= - \frac{\alpha \dot{M}_{\footnotesize \mbox{star}}}{2}\left(\frac{R_p}{D}\right)^2.
\end{align}
Following \cite{rami16} we use an entrainment efficiency of $\alpha$ = 0.2 for our Earth-like atmospheres, assuming an Earth-mass planet with a 1~bar surface pressure.  Substituting in $\dot{M}_{star}$ from Equations~\ref{RG_m} and~\ref{AG_m} we calculate planetary atmospheric loss throughout the post-MS.

\subsection{Post-MS stellar model spectra}

The stellar input spectra consist of a combination of observed data in the UV (where models poorly represent stellar irradiation) from the \emph{International Ultraviolet Explorer} (\emph{IUE})\footnote{http://archive.stsci.edu/iue}, an astronomical observatory satellite primarily designed to take UV spectra.  We use type III luminosity class stars drawn from \cite{lope05}, \cite{luck15}, and \cite{stoc18} with available long wave (LW) and short wave (SW) IUE spectra from 1216 to 3347~\AA\ in combination with synthetic spectra from the Pickles Atlas \citep{pick98} from 3347 to 45450~\AA\ following \cite{rugh13}.  Our star sample is drawn primarily from close-by post-MS stars but includes more distant targets to allow for a more complete sample of observed IUE data for a range of post-MS stars with different spectral types, predicted masses, and evolutionary phases.  

\begin{table}[h!]
\caption{IUE spectra hosts \label{RG_targets}}
\begin{center}
\small
\begin{tabular}{cclrrrrrrrl}
\hline
Mass  & Spectral & Star & Star & Model & Dist. & Radius & No.\ SW& No.\ LW & Scale & Evolution \\
 (\Ms) & Type & Name & $T_{\footnotesize \mbox{eff}}$ (K) & $T_{\footnotesize \mbox{eff}}$ (K) & (pc) & (R$_\odot$) & spectra & spectra & factor & stage \\
 \hline
3.0	& G5 III	& HD 74772 			& 5118 	& 5164 &70.18		& 12.90	& 2	& 2	& 2.0		& HB		\\ 
3.0	& G8 III	& HD 148374 			& 4948 	& 5011 & 155.28 	& 14.52	& 2	& 1	& 2.2		& HB*	\\ 
2.3	& K0 III	& $\beta$ Gem 			& 4865 	& 4853 &10.36		& 8.8		& 12	& 9	& 2.4		& HB		\\ 
3.0	& K0 III	& $\beta$ Ceti 			& 4797 	& 4853 & 29.5 		& 16.78	& 5	& 4	& 2.4		& HB		\\ 
1.3	& K2 III	& $\iota$ Draconis 		& 4445 	& 4457 & 31.03 	& 11.99	& 1	& 2	& 2.7		& HB		\\ 
1.3	& K2 III	& $\theta$ Doradus 		& 4320 	& 4457 & 151 		& 16		& 1	& 1	& 1.5		& HB/AGB*	\\ 
2.0	& K3 III	& $\alpha$ Boo 		& 4286 	& 4365 & 11.26 	& 25.4	& 29	& 10	& 1.4		& RGB	\\ 
2.3	& K5 III	& $\gamma$ Draconis 	& 3989 	& 4009 & 47.3 		& 53.4	& 4	& 2	& 2.2		& HB/AGB*	\\ 
2.0	& M5 III	& $\gamma$ Cru 		& 3626 	&  3819 & 27.2 		& 84		& 28	& 11	& 2.2		& AGB*	\\
 \hline
 \end{tabular}
 *Indicates that the target's evolutionary stage was determined by H-R diagram fitting
 \end{center}
 \end{table}

\begin{figure}[b!]
\centering
\includegraphics[scale=0.4]{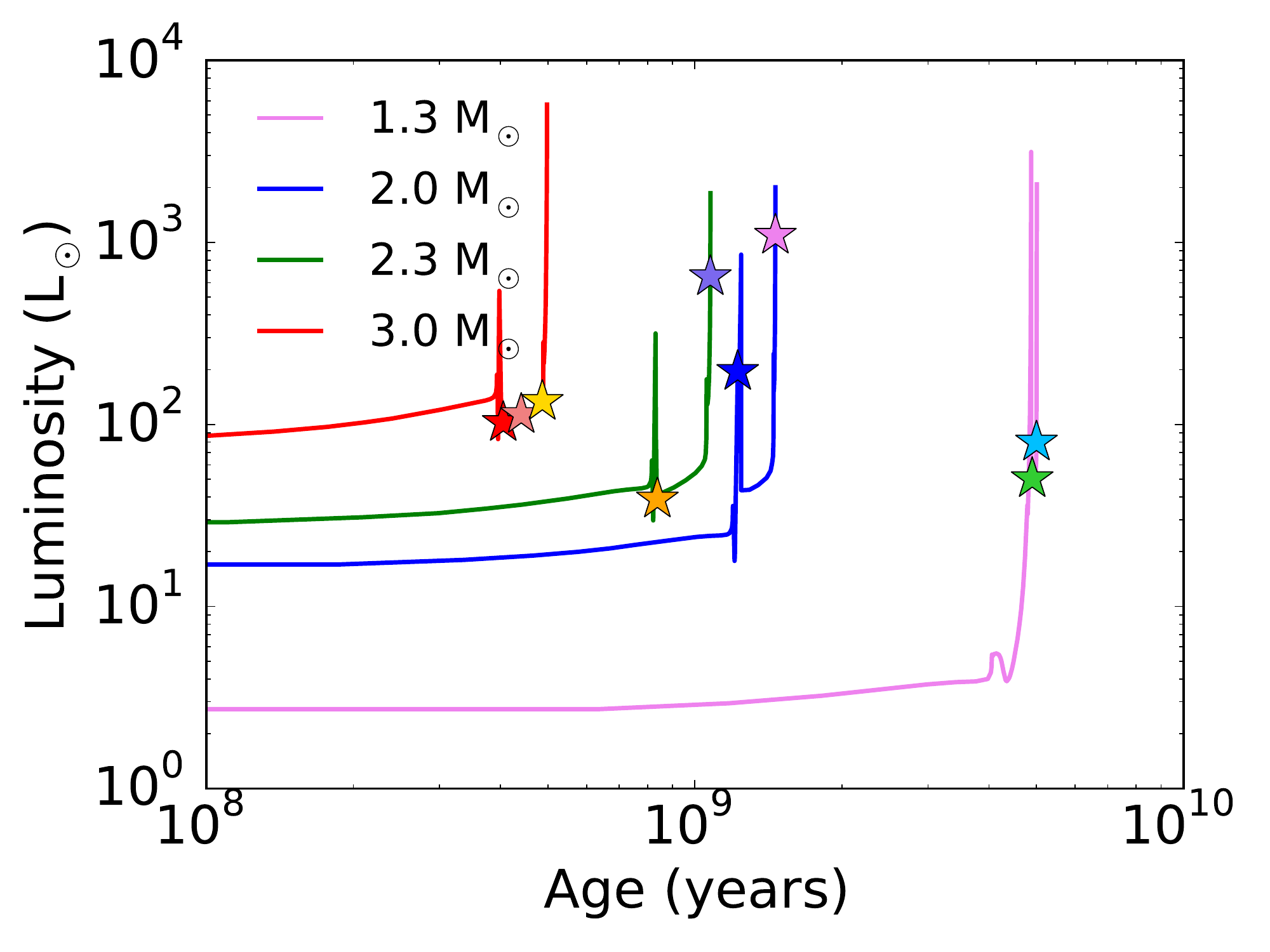}
\includegraphics[scale=0.4]{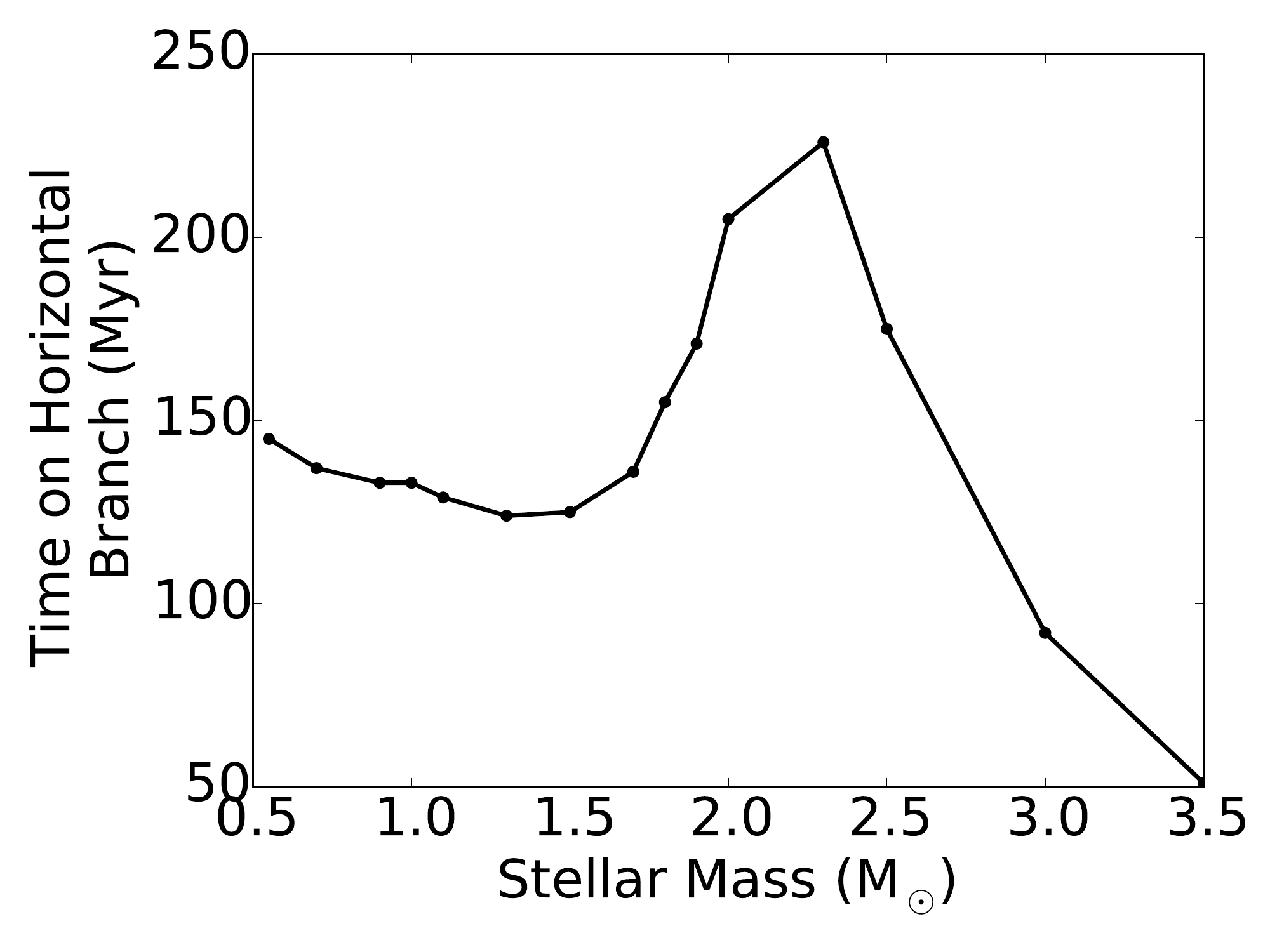}
\caption{(Left) Post-MS IUE targets (marked as stars) fit to the Padova evolutionary tracks and (right) the length of the HB for each specific mass track.   More information on targets shown in Table~\ref{RG_targets}.  \label{HR_fitting}}
\end{figure}

We predict the IUE target's original mass using H-R diagram fitting with the Padova evolutionary tracks and, if available, also use the predicted evolutionary phase from \cite{stoc18}. For stars without an estimated evolutionary phase, we use the predicted age from HR-diagram fitting (see Figure~\ref{HR_fitting}). Table~\ref{RG_targets} lists stellar data along with the number of SW (1150 to 1979~\AA) and LW (1979 to 3347~\AA) IUE spectra median-combined for each target, as well as the scale factor applied to match up the combined spectrum to the corresponding Pickles Atlas synthetic spectrum.

\begin{figure}[h!]
\includegraphics[scale=0.4]{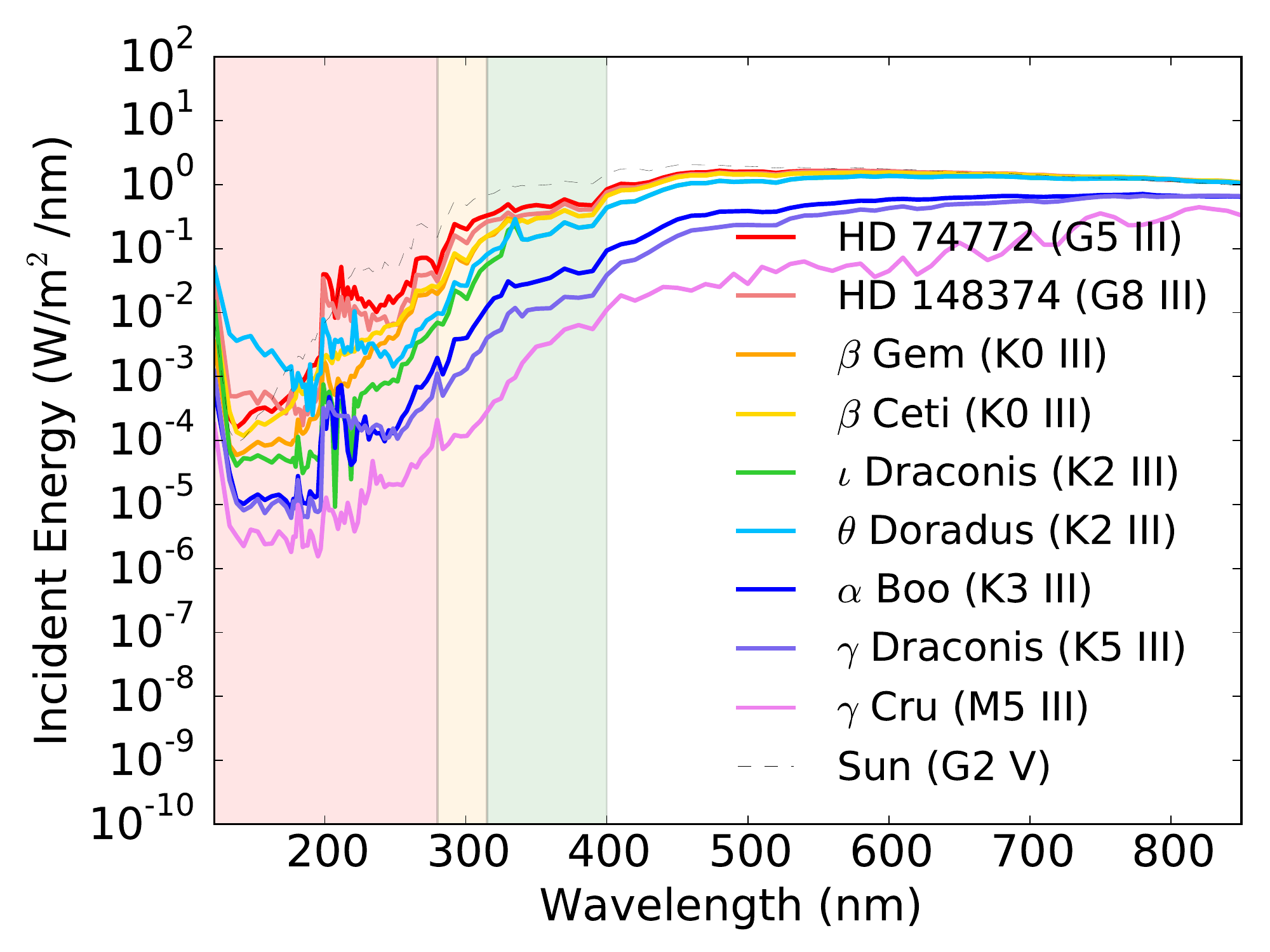}
\includegraphics[scale=0.4]{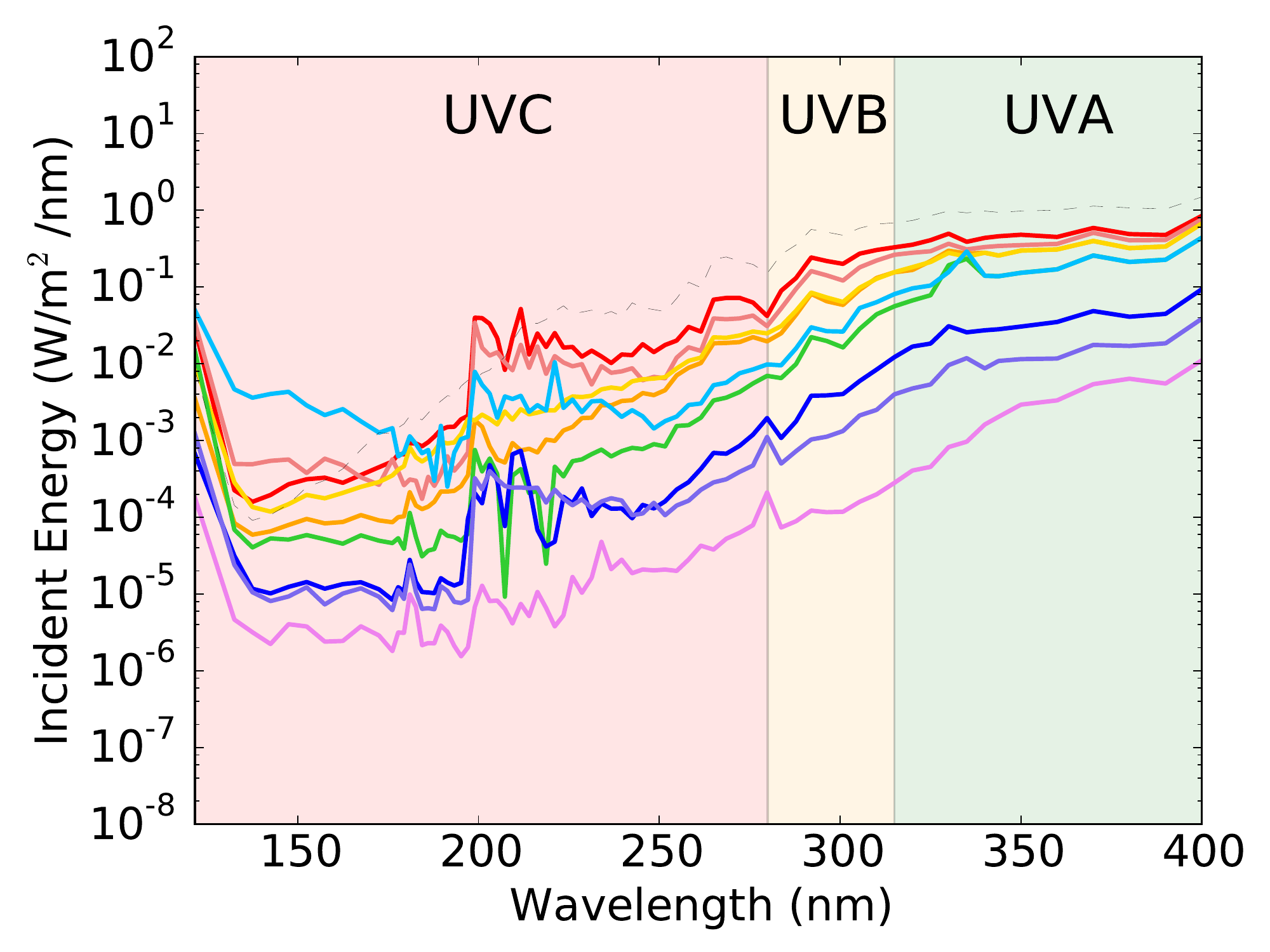}
\caption{Red giant spectra scaled to the Earth-irradiation based on a combination of IUE data and Pickles model spectra for our grid stars (Table~\ref{RG_targets}). The UV irradiation present-day Earth receives from the Sun (G2 V) is shown as a black dotted line. \label{RG_models}}
\end{figure}

Figure~\ref{RG_models} shows the spectra for the nine post-MS stars in Table~\ref{RG_targets} for a 1~AU equivalent distance from the star, where a planet would receive the same irradiation as present-day Earth.

\subsection{Modeling planetary atmospheres and UV surface environments}

We use EXO-Prime (see, e.g.\ \cite{kalt10}), a coupled 1D radiative-convective atmosphere code based on iterations of a 1D climate model \citep{kast86,pavl00,haqq08} and 1D photochemistry model \citep{pavl02,segu05,segu07} developed for rocky exoplanets.  Models simulate the effects of stellar and planetary conditions on exoplanet atmospheres and surface environments and are run to convergence following \cite{segu05}.  Model atmospheres extend to an altitude of 60 km (pressure of 1~mbar) divided up into 100 parallel planes using a stellar zenith angle of 60$^\circ$.  LW IR fluxes are calculated with a rapid radiative transfer model, and SW visible/near-IR fluxes are calculated with a two-stream approximation following \cite{toon89} with atmospheric gas scattering.  The photochemistry code includes 55 chemical species with 220 reactions solved with a reverse-Euler method originally developed by \cite{kast85}.

We scale the incident top-of-the-atmosphere (TOA) stellar flux for our planet models to match the total integrated flux of the Sun-Earth system to compare the effects of post-MS stellar irradiance to present-day Earth.  Planetary outgassing rates are kept constant for H$_2$, CH$_4$, CO, N$_2$O, and CH$_3$Cl and maintain constant mixing ratios of O$_2$ at 0.21 and CO$_2$ at 3.55$\times10^{-6}$, while N$_2$ concentrations vary to reach the initial surface pressure model condition (following  \cite{segu03,segu05,rugh13,rugh15,rugh15b,rugh18,koza18}).

Many reactions in Earth's atmosphere are driven by the Sun's UV flux. Some atmospheric species exhibit noticeable features in our planet's spectrum as a result directly or indirectly of biological activity. Oxygen or ozone in combination with a reducing gas like methane are currently our best biosignatures, indicating biological activity on a temperate rocky planet (see review \cite{kalt17}). Ozone (O$_3$) is created and destroyed with UV photons through the Chapman reactions  \citep{chap30}, 
\begin{equation}
\begin{aligned}
\m{O}_2 + \m{h}\nu \rightarrow \m{O + O } (\lambda < 240\ \m{nm}), \\
\m{O + O}_2 + M \rightarrow \m{O}_3 + M, \\
\m{O}_3 + \m{h}\nu \rightarrow \m{O}_2 \m{ + O}\ (\lambda < 320\ \m{nm}),\\
\m{O}_3 + \m{O} \rightarrow \m{2O}_2, 
\end{aligned}
\label{O3}
\end{equation}
where $M$ is a background molecule such as N$_2$.  These reactions are primarily responsible for ozone production on present-day Earth and are dependent on the UV portion of the stellar host's spectrum.

In addition, atmospheric ozone levels determine the amount of UV flux that reaches the planetary surface on present-day Earth (see, e.g.\ \cite{rugh15,omal17,koza18}), which can impact surface life. We focus here in particular on UVB (280-315~nm) and UVC (100-280~nm). UVB irradiation, which can be damaging for surface life, is partially shielded by ozone, whereas UVC irradiation, which is energetic enough to cause DNA damage, is almost entirely shielded by ozone, along with being the primary source that creates ozone. We explore the UV environment on the surface of our model planets following \cite{rugh15}, \cite{omal17}, and \cite{koza18} and compare it to present-day Earth's.

\section{Results \label{results}}

\subsection{Orbital distance of the post-MS HZ \label{HZ_dist}}

We modeled the orbital distance of the post-MS HZ for host stars of~1.0 to~3.5~\Ms. Then we  calculated the initial semimajor axis required for a planet to spend the maximum amount of time in the post-MS HZ while undergoing semimajor axis evolution (see Equation~\ref{D_t}). Post-MS stars that undergo a helium flash (0.8~to~2~\Ms) sustain significantly larger luminosity increases (factors of order~1000) when compared to higher-mass stars that do not experience helium flashes (factors of order~100). As a result, planets around lower-mass stars are subject to larger amounts of semimajor axis variation due to higher stellar mass-loss rates.

Table~\ref{HZ_limits} shows the factor by which stellar luminosity increases on the post-MS,  the post-MS HZ boundaries for the beginning and end points of the HB (the most stable part of the post-MS), and and the semimajor axis evolution of the orbit of a planet that could spend the maximum possible time in the post-MS HZ.  Post-MS HZ orbital distances are comparable for lower-mass stars (1.0 and 1.3~\Ms) and higher-mass stars (2.0 and 2.3~\Ms) due to the much higher luminosity increase for our grid stars that undergo helium flashes ($<$~2~\Ms).  Greater luminosity increases during the post-MS additionally cause higher stellar mass-loss rates (Equations~\ref{RG_m} and~\ref{AG_m}), which, in turn, cause larger variations in an orbiting planet's semimajor axis for our lower-mass grid stars (Equation~\ref{D_t}).

\begin{table}[h!]
\begin{center}
\small
\caption{HZ boundaries and semimajor axis evolution \label{HZ_limits}}
\begin{tabular}{cr|rr|rr|rrr }
\hline
Track Mass & $L$ factor & \multicolumn{2}{c}{ HB start (AU)} & \multicolumn{2}{c}{ HB end (AU)} &\multicolumn{3}{c}{Semimajor axis* (AU)}  \\
(\Ms) & increase & Inner HZ &	Outer HZ & Inner HZ & Outer HZ &  Initial & HB start & HB end    \\
\hline	
1.0	&	4457 &	5.1	&	12.7 	&	5.6	&	13.9 &	10.0	&	12.3 &	12.3  \\
1.3	&	1258 &	5.6	&	14.0	&	12.0	&	31.0 &	12.5	&	13.8 &	13.9  \\
2.0	&	126 &	5.2	&	12.6	&	12.2	&	31.0	&	12.2	&	12.3	&	12.3	\\
2.3	&	68 &		5.0	&	12.0	&	10.5	&	26.4	&	12.0 &	12.0	&	12.0 \\
3.0	&	74 &		7.5	&	18.3	&	13.3	&	33.3	&	18.2	&	18.2	&	18.2	\\
3.5	& 	73 &	10.5	&	25.6	&	14.7	&	36.7	&	25.0	&	25.0	&	25.0	\\
\hline
\end{tabular}
\end{center}
*Semimajor axis values correspond to the orbit that spends the maximum amount of time in the post-MS HZ
\end{table}

\subsection{Post-MS habitable zone lifetime}

\begin{figure}[tp]
\begin{center}
\includegraphics[scale=0.37]{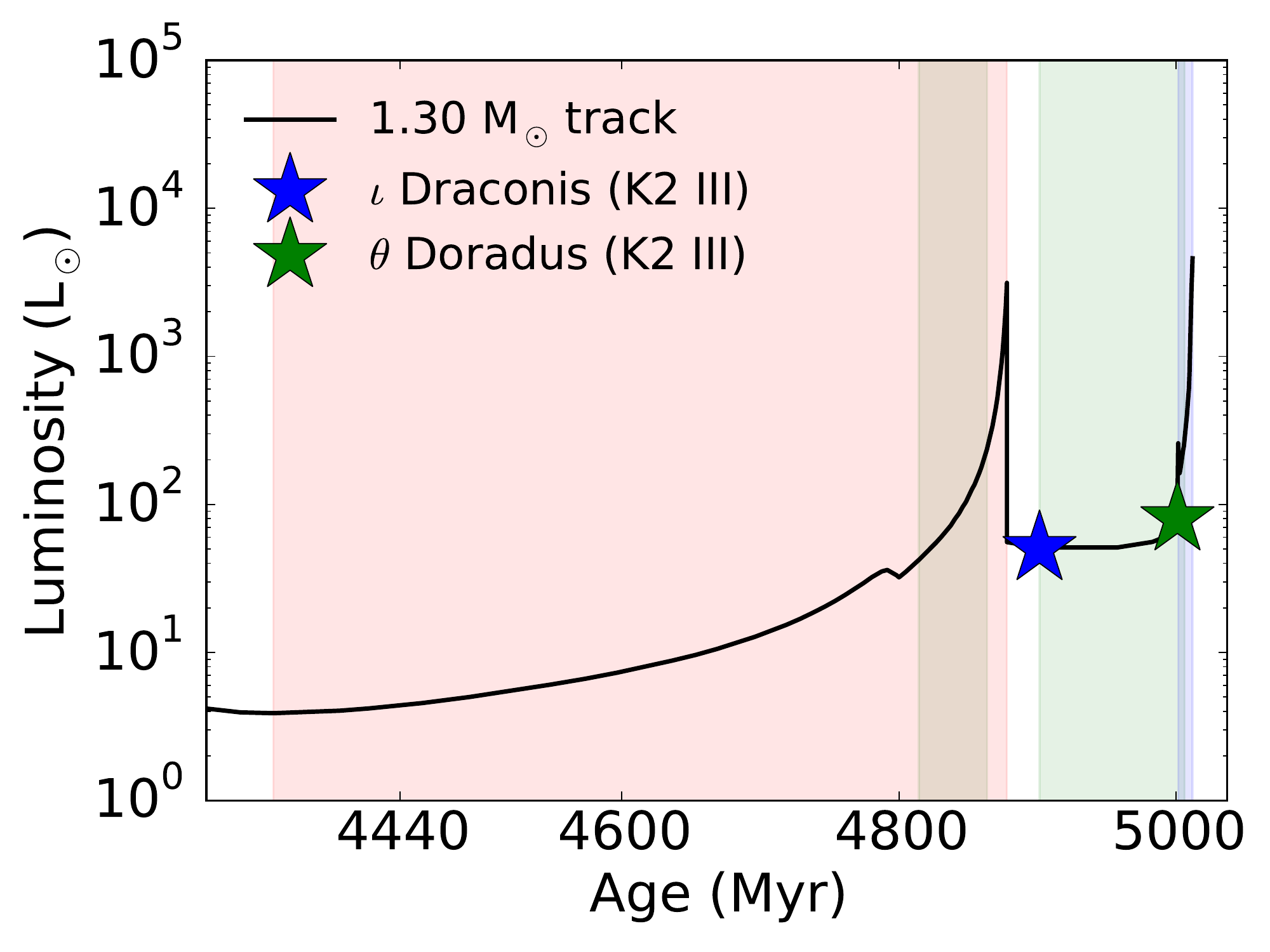}
\includegraphics[scale=0.37]{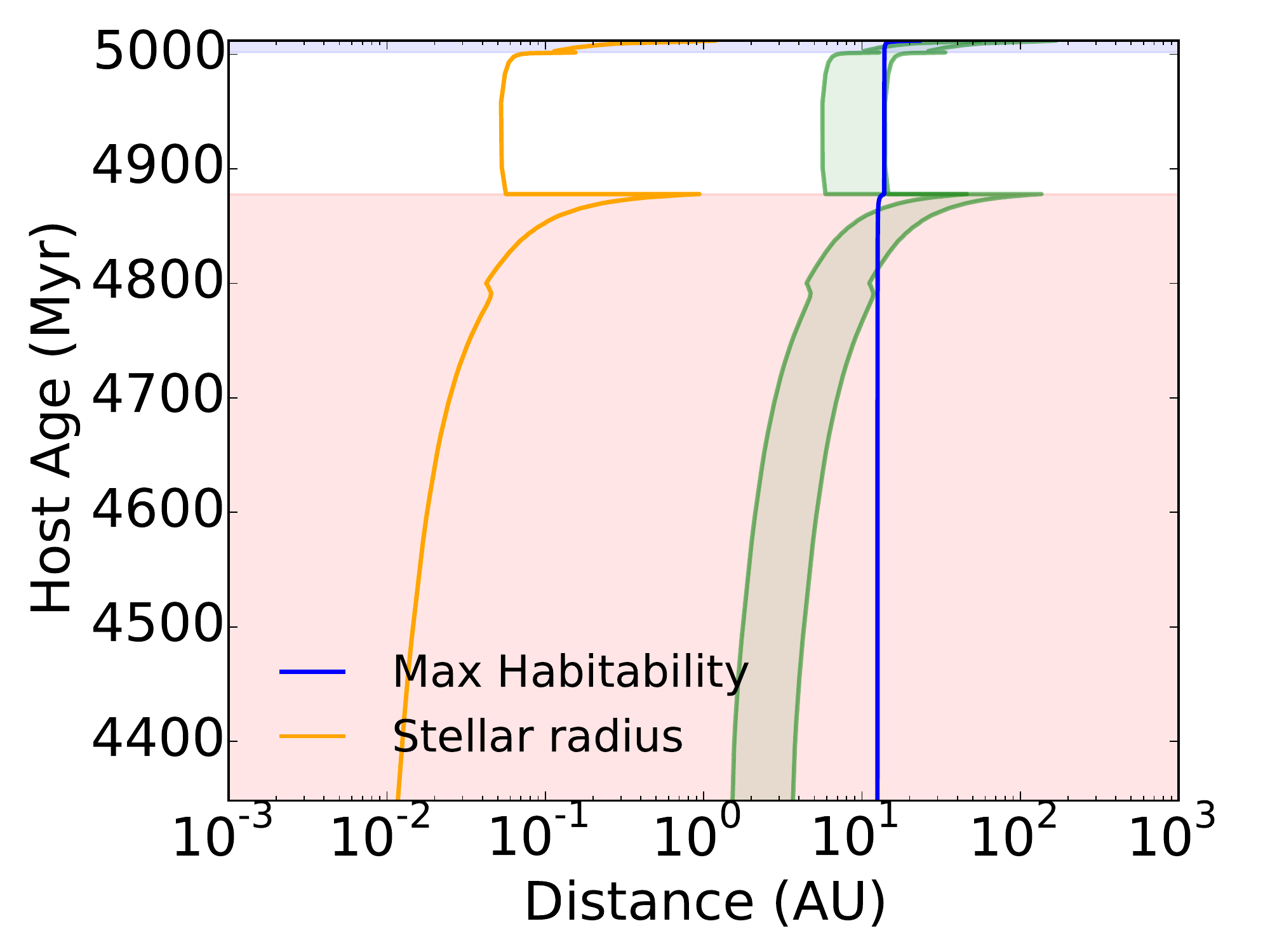} \\
\includegraphics[scale=0.37]{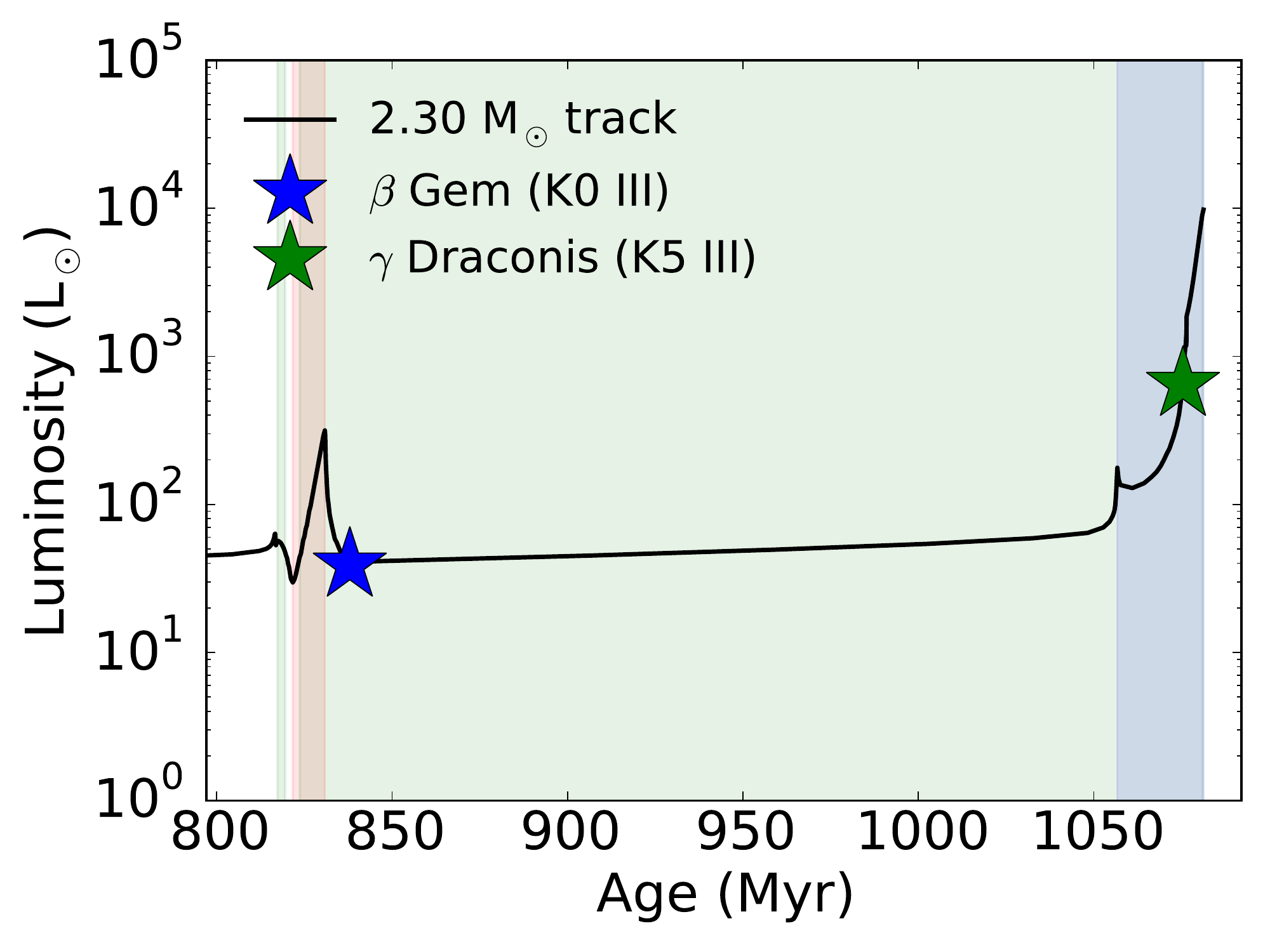}
\includegraphics[scale=0.37]{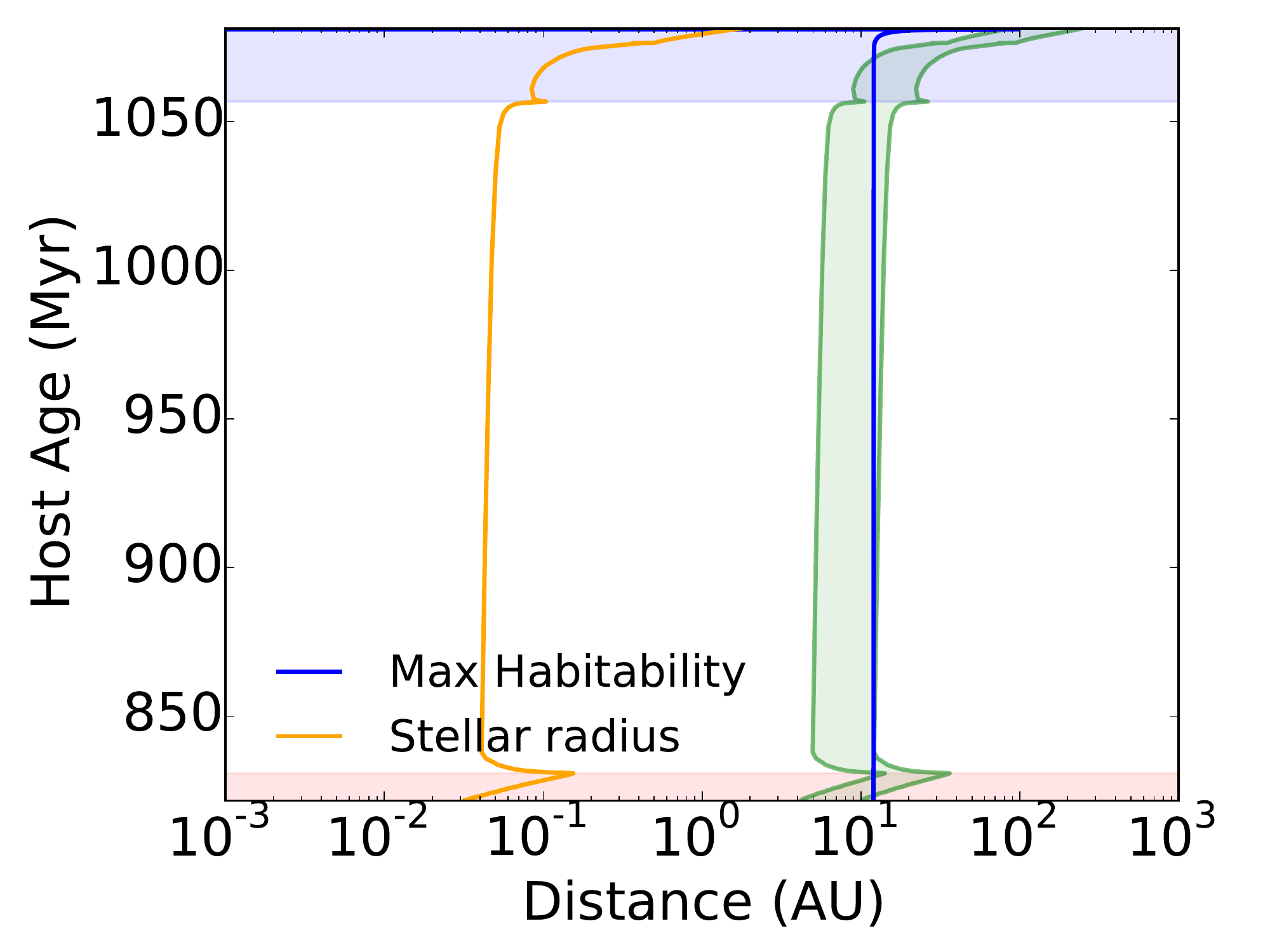} \\
\includegraphics[scale=0.37]{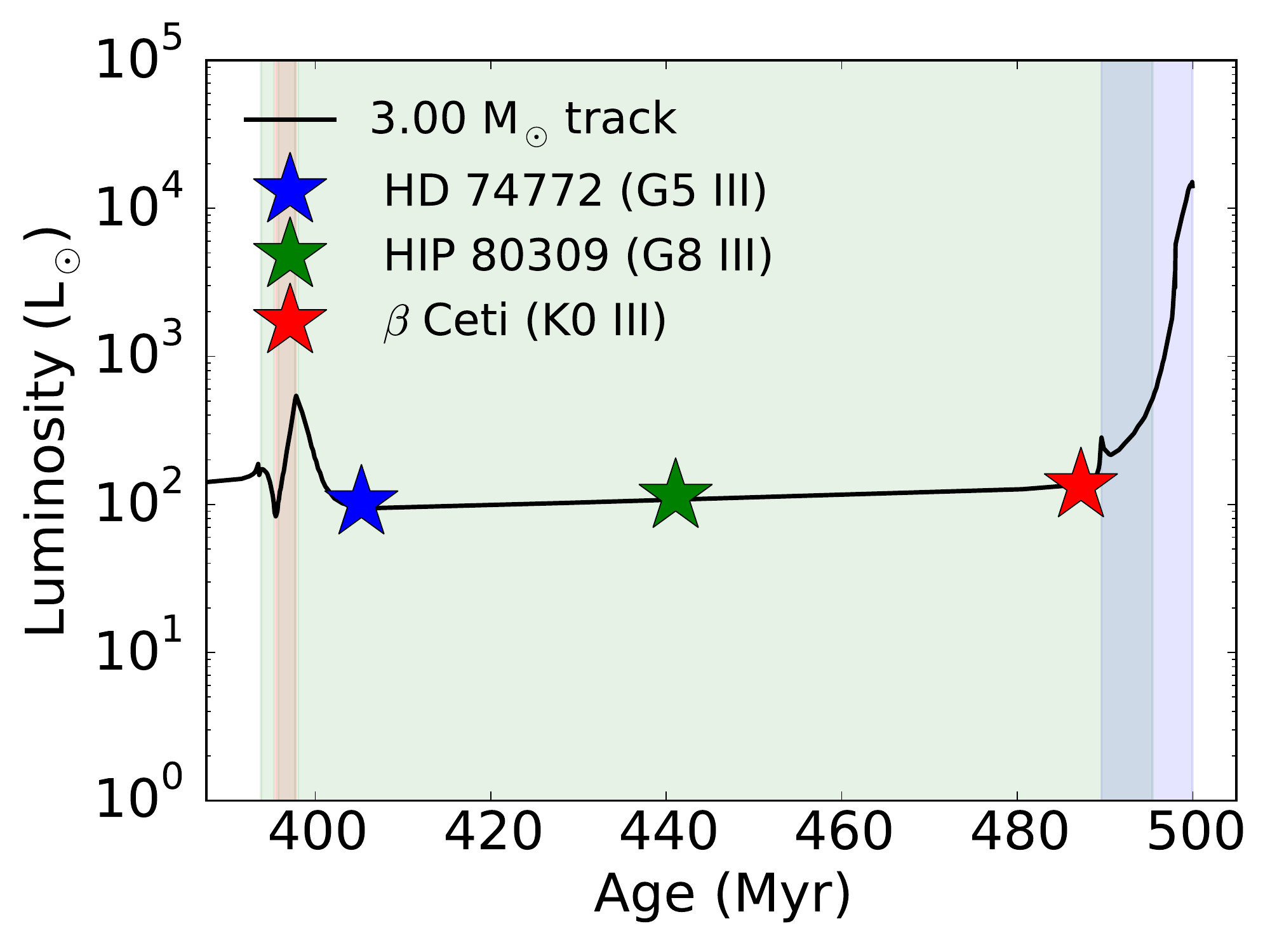}
\includegraphics[scale=0.37]{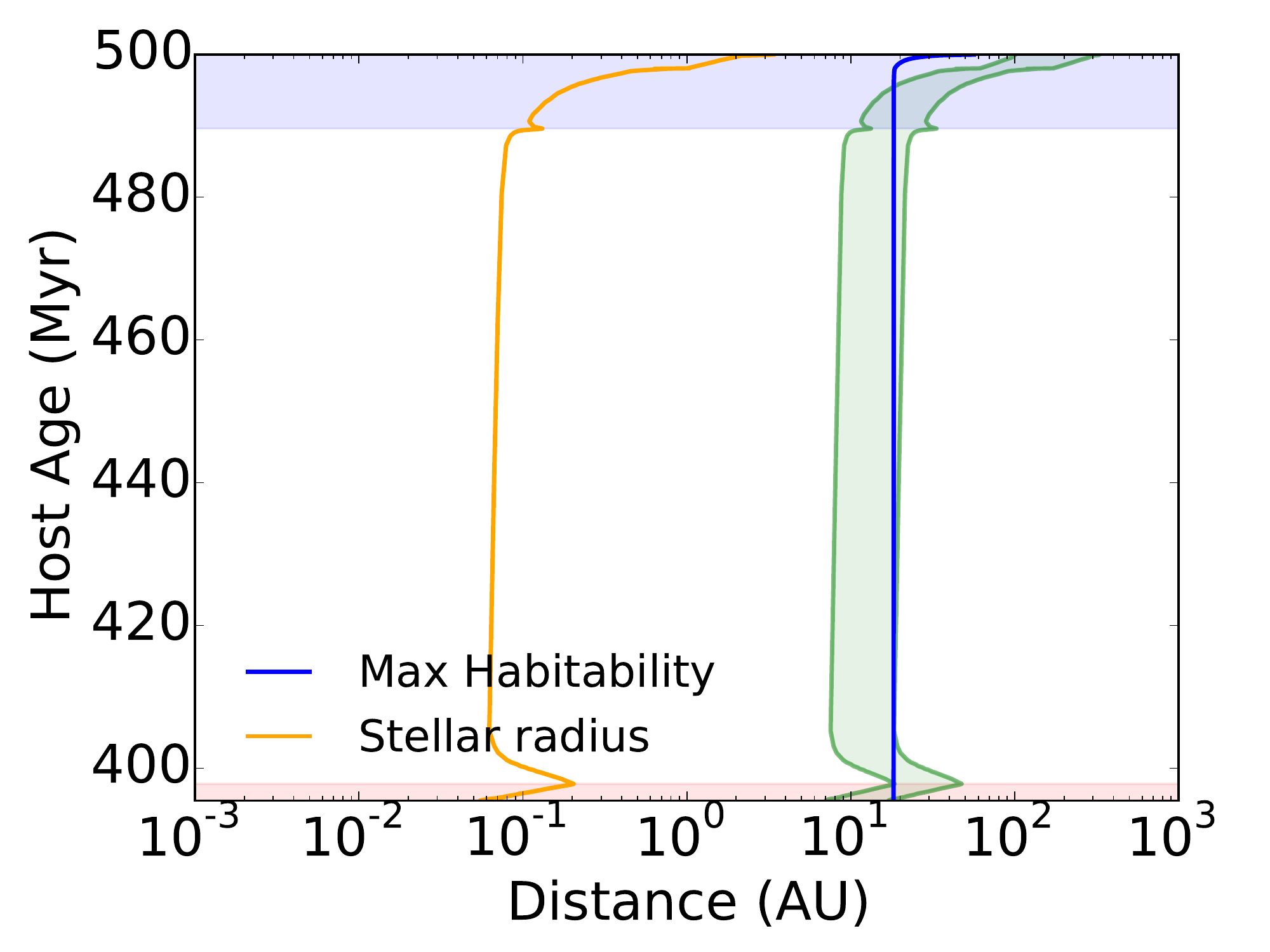}
\end{center}
\caption{(Left) Stellar luminosity versus stellar age during the post-MS with grid stars indicated on the evolutionary track with stars, and (right) evolution of the orbital distances of the post-MS HZ for (top) a 1.3, (middle) a 2.3, and (bottom) a 3.0~\Ms\ track star. For the plots on the left the red zone indicates the RGB, the blue zone the AGB, and the green zone the time that the model planet passes through the HZ. The points of evolution modeled in Section~\ref{evol_sec} are marked with stars.  For plots on the right the green zone indicates the changing orbital distance of the HZ, the orange line the evolution of the host star's radius, and the blue line the orbital evolution of an Earth-sized planet spending the maximum amount of time in the HZ.  \label{HZ_tracks}}
\end{figure}

For the modeled post-MS HZs and orbital evolution of 1~Earth-mass planets for stars of masses~1.0 to 3.5~\Ms, we explore the longest time a planet can orbit within the post-MS HZ. We also estimate the resulting planetary atmosphere erosion due to the host star's mass loss (see Table~\ref{HZ_time}). Note that the longest time spent in the post-MS HZ does not need to be continuous due to the host stars changing luminosity in the post-MS phase. Thus, we distinguish between the time an Earth-mass planet can spend continuously in the post-MS HZ (post-MS CHZ), and the maximum time a planet can spend in the post-MS HZ that is not continuous and can put the planet in a  temporary runaway greenhouse state due to the large increase of stellar luminosity during the peak of the RGB (see Figure~\ref{HZ_tracks} and discussion).

A star with a ZAMS mass of 1.0~\Ms\ has a total lifetime of 12.7~Gyr, spending only 8\% of its life on the post-MS with 851 Myr on the RGB, 133~Myr on the HB, and 27~Myr on the AGB. The initial semimajor axis for a planet leading to the maximal time in the post-MS HZ is 10.0~AU. It spends 66~Myr in the post-MS HZ on the RGB, 22~Myr outside of the post-MS HZ during the RGB peak, where stellar irradiation exceeds the empirical HZ limits, and then remains in the post-MS HZ for 153~Myr on the HB. The stellar irradiation exceeds the empirical HZ limits again during the AGB. Stellar winds erode about 10\% of the planet's original atmosphere when its star's irradiation exceeds the empirical HZ limits during the AGB.

A star with a ZAMS mass of 1.3~\Ms\ has a total lifetime of 5.0 Gyr, spending 13\% of its life on the post-MS with 529~Myr on the RGB, 124~Myr on the HB, and 10~Myr on the AGB. The initial semimajor axis for a planet leading to the maximum amount of time in the post-MS HZ is 12.5~AU. It spends 49~Myr in the post-MS HZ on the RGB, 38~Myr outside of the post-MS HZ during the RGB peak, where stellar irradiation exceeds the empirical HZ limits, and then remains in the post-MS HZ for 105~Myr on the HB and the beginning of the AGB. Stellar winds erode about 5\% of the planet's original atmosphere when its star's irradiation exceeds the empirical HZ limits during the AGB.

\begin{table}[h!]
\begin{center}
\small
\caption{Post-MS HZ lifetime \label{HZ_time}}
\begin{tabular}{crrrrr}
\hline
Track Mass & Total post- & Max HZ & Max CHZ  &Time outside & Atmosphere  \\
(\Ms)  & MS (Myr) & (Myr) & (Myr) & HZ$^1$ (Myr) & Eroded (\%)  \\
\hline
1.0	&	1011	&	219 &	153	&	22	&	10  \\
1.3	&	663	&	154 &	105	&	38	&	5  \\
2.0	&	260	&	191 &	163	&	50	&	1  \\
2.3	&	260	&	259 &	257	&	4	&	0.1  \\
3.0	&	104	&	101 &	97	&	1	&	0.01  \\
3.5	&	58	&	56 &		56	&	0	&	0.1 \\		
\hline
\end{tabular}
\end{center}
*Planetary parameters are those of the orbit of maximum time in the HZ \\
$^1$ Time spent outside of the HZ during the RGB after initially entering HZ
\end{table}

A star with a ZAMS mass of 2.0~\Ms\ has a total lifetime of 1.5~Gyr, spending 18\% of its life on the post-MS with 38~Myr on the RGB, 206~Myr on the HB, and 16~Myr on the AGB. The initial semimajor axis for a planet leading to the maximum amount of time in the post-MS HZ is 12.2~AU. It spends 28~Myr  in the post-MS HZ on the RGB, 50~Myr outside of the post-MS HZ during the RGB peak, where stellar irradiation exceeds the empirical HZ limits, and then remains in the post-MS HZ for 163~Myr on the HB. Stellar winds erode about 1\% of the planet's original atmosphere when its star's irradiation exceeds the empirical HZ limits during the AGB.

A star with a ZAMS mass of 2.3~\Ms\ has a total lifetime of 1.08~Gyr, spending 24\% of its life on the post-MS with 9~Myr on the RGB, 226~Myr on the HB, and 25~Myr on the AGB. The initial semimajor axis for a planet leading to the maximum amount of time in the post-MS HZ is 12.0~AU. It spends 2~Myr in the post-MS HZ on the RGB, 4~Myr outside of the post-MS HZ during the RGB peak, where stellar irradiation exceeds the empirical HZ limits, and then remains in the post-MS HZ for 257~Myr on the HB and the beginning of the AGB. Stellar winds only erode 0.1\% of the planet's original atmosphere when its star's irradiation exceeds the empirical HZ limits during the AGB.

A star with a ZAMS mass of 3.0~\Ms\ has a total lifetime of 500~Myr, spending 21\% of its life on the post-MS with 2~Myr on the RGB, 92~Myr on the HB, and 6~Myr on the AGB. The initial semimajor axis for a planet leading to the maximum amount of time in the post-MS HZ is 18.2~AU.  The planet spends a total of 4~Myr in the post-MS HZ on the RGB, 1~Myr outside of the post-MS HZ when stellar irradiation exceeds the empirical HZ limits during the RGB peak, and then remains in the post-MS HZ for 97~Myr on the HB and the beginning of the AGB.  Stellar winds only erode 0.01\% of the planet's original atmosphere when its stars irradiation exceeds the empirical HZ limits during the AGB, due to the wide orbital separation and the brevity of the star's lifetime in our model.

A star with a ZAMS mass of 3.5~\Ms\ has a total lifetime of 321~Myr, spending 18\% of its life on the post-MS with 1~Myr on the RGB, 51~Myr on the HB, and 6~Myr on the AGB. Due to a smaller overall luminosity change for stars this massive, a continuous post-MS HZ (CHZ) for a planet exists. The initial semimajor axis for a planet leading to the maximum amount of time of 56~Myr in the post-MS HZ is 25~AU from the RGB to the AGB. Stellar winds erode about 0.1\% of the planet's original atmosphere when its star's irradiation exceeds the empirical HZ limits during the AGB, despite the large semimajor axis because of prolonged exposure to high stellar mass-loss rates on the AGB.

\subsection{Planets at Earth-equivalent orbital distances \label{earth-like}}

\begin{figure}[h!]
\centering
\includegraphics[scale=0.45]{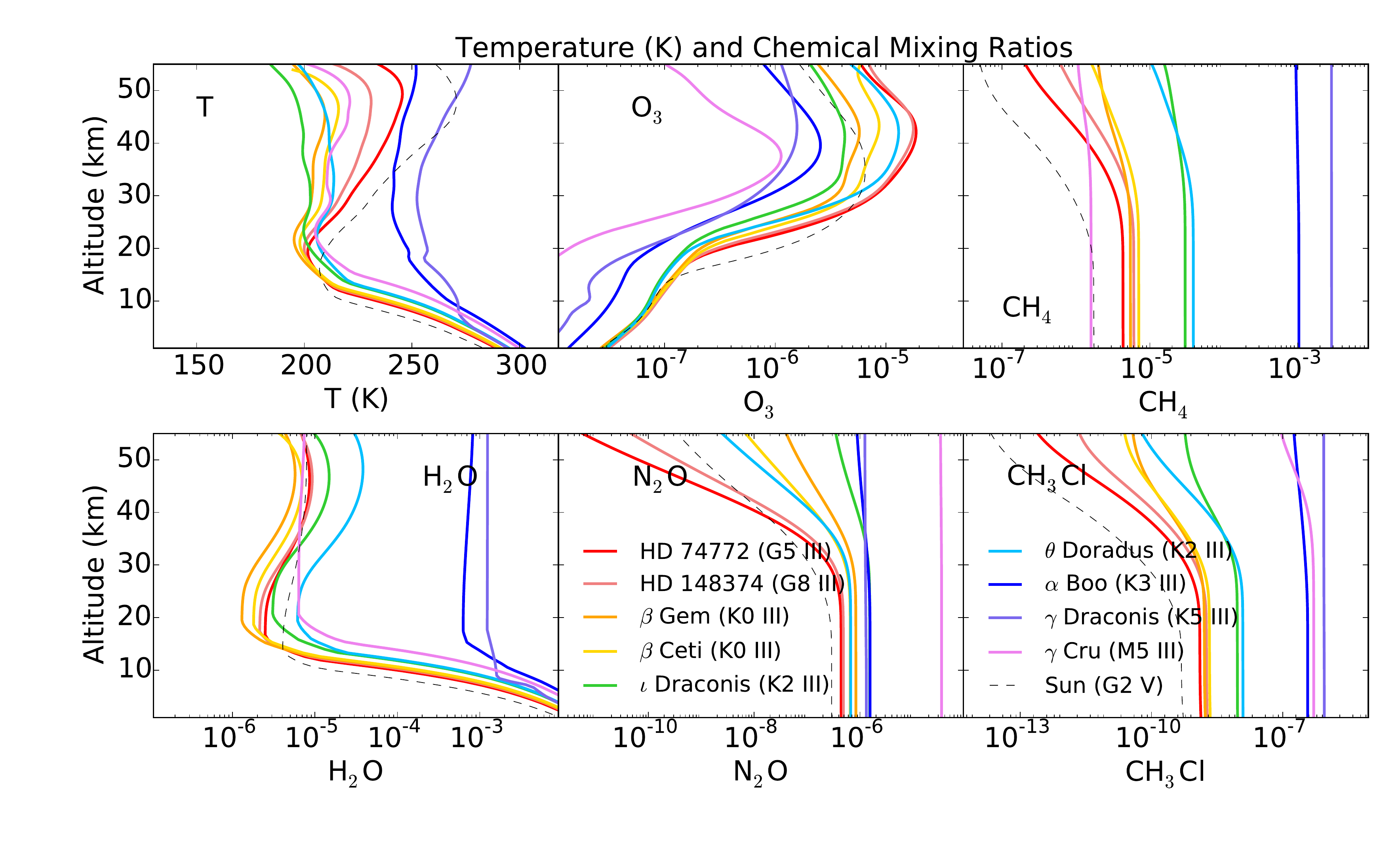}
\caption{Temperature and atmospheric mixing ratio profiles for an Earth-like planet orbiting our grid post-MS stars at the Earth-equivalent distance. \label{profiles}}
\end{figure}

We modeled atmospheres of Earth-like planets at distances where they receive the same integrated flux as present-day Earth ($S_{\footnotesize \mbox{eff}}$~=~1) with surface pressures of 1.0~bar. We focus on the change of atmospheric signatures that can indicate life on a planet, ozone and oxygen in combination with a reducing gas like methane or N$_2$O, as well as climate indicators such as water and CO$_2$, which can also indicate whether the oxygen production can be explained abiotically (see review \cite{kalt17}). We also explore the  amount of UV flux reaching the model planetary surface in the post-MS HZ throughout the host star's evolution later in Section~\ref{evol_sec}. The temperature and mixing ratios of the resulting model atmospheres are shown for all planet models for our our grid stars in Figure~\ref{profiles}, with present-day Earth's profiles shown as black dashed lines for comparison.

As a star expands and cools  during the post-MS phase, the peak of the stellar spectrum shifts to redder wavelengths, which heat planetary surfaces more efficiently for Earth-like planets with a mostly N$_2$-H$_2$O-CO$_2$ atmosphere (see e.g.\ \cite{kast93}). This is partly due to the effectiveness of Rayleigh scattering, which decreases at longer wavelengths. A second effect is the increase in NIR absorption by H$_2$O and CO$_2$ as the star's spectral peak shifts to these wavelengths, meaning that the same integrated stellar flux that hits the top of a planet's atmosphere from a cool red star warms a planet more efficiently than the same integrated flux from a star with a higher effective surface temperature. This causes planetary surface temperatures to be lower for planets orbiting hotter stars than cooler stars, even with the same amount of total integrated incident flux, as shown in Figure~\ref{profiles} and Table~\ref{model_summary}. 

Note that due to the drop in stellar temperature during expansion on the RGB, the average surface temperature of a post-MS star is lower than for MS stars with similar masses. All grid stars in our study have lower effective temperatures than the Sun, and as a result, all calculated surface temperatures for planets receiving Earth-like irradiation are higher than present-day Earth's (shown as a dashed black line), ranging from 1.1~K for the hottest to 18.2~K higher for the coolest post-MS grid star.

\begin{figure}[h!]
\centering
\includegraphics[scale=0.4]{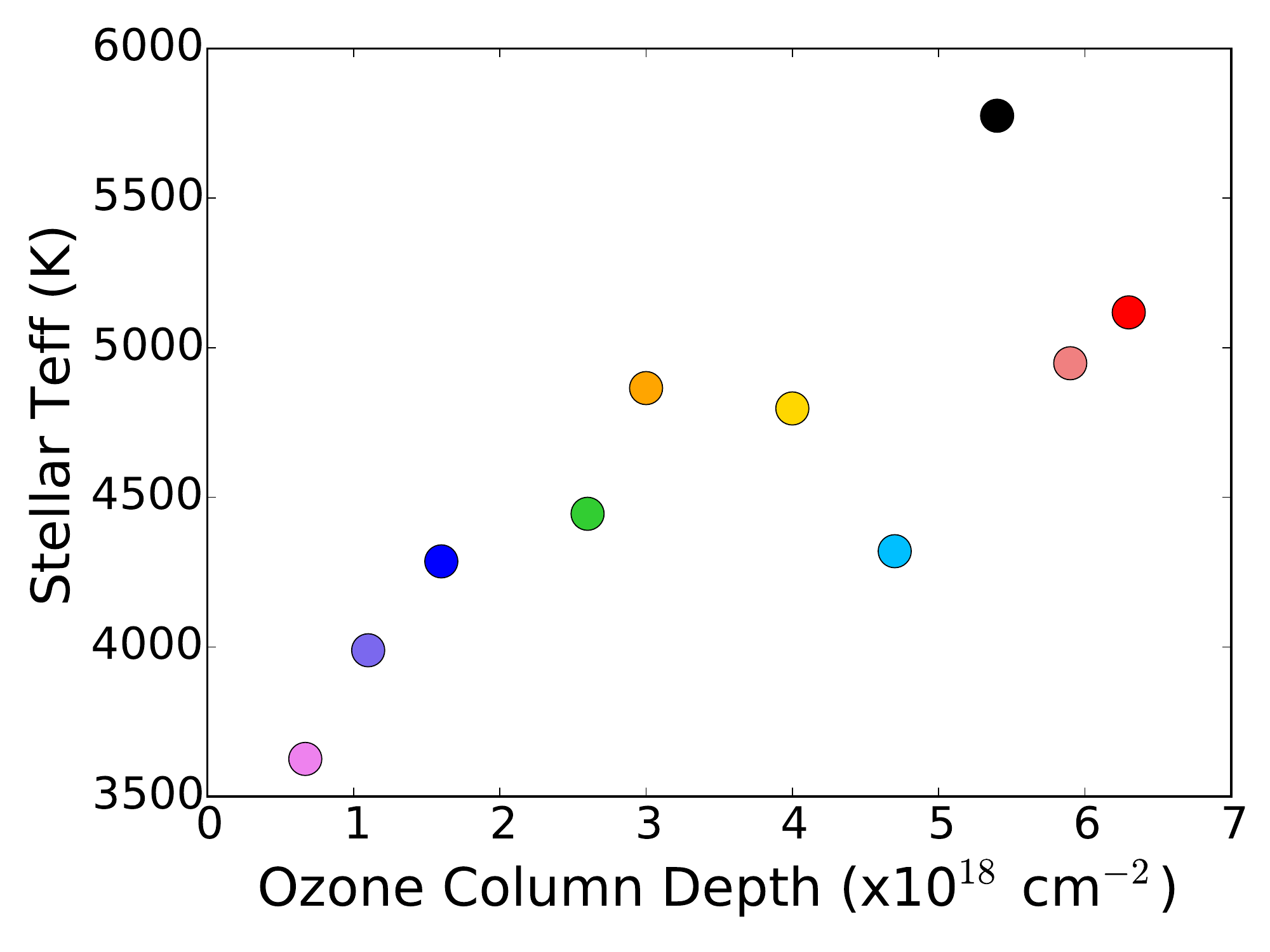}
\includegraphics[scale=0.4]{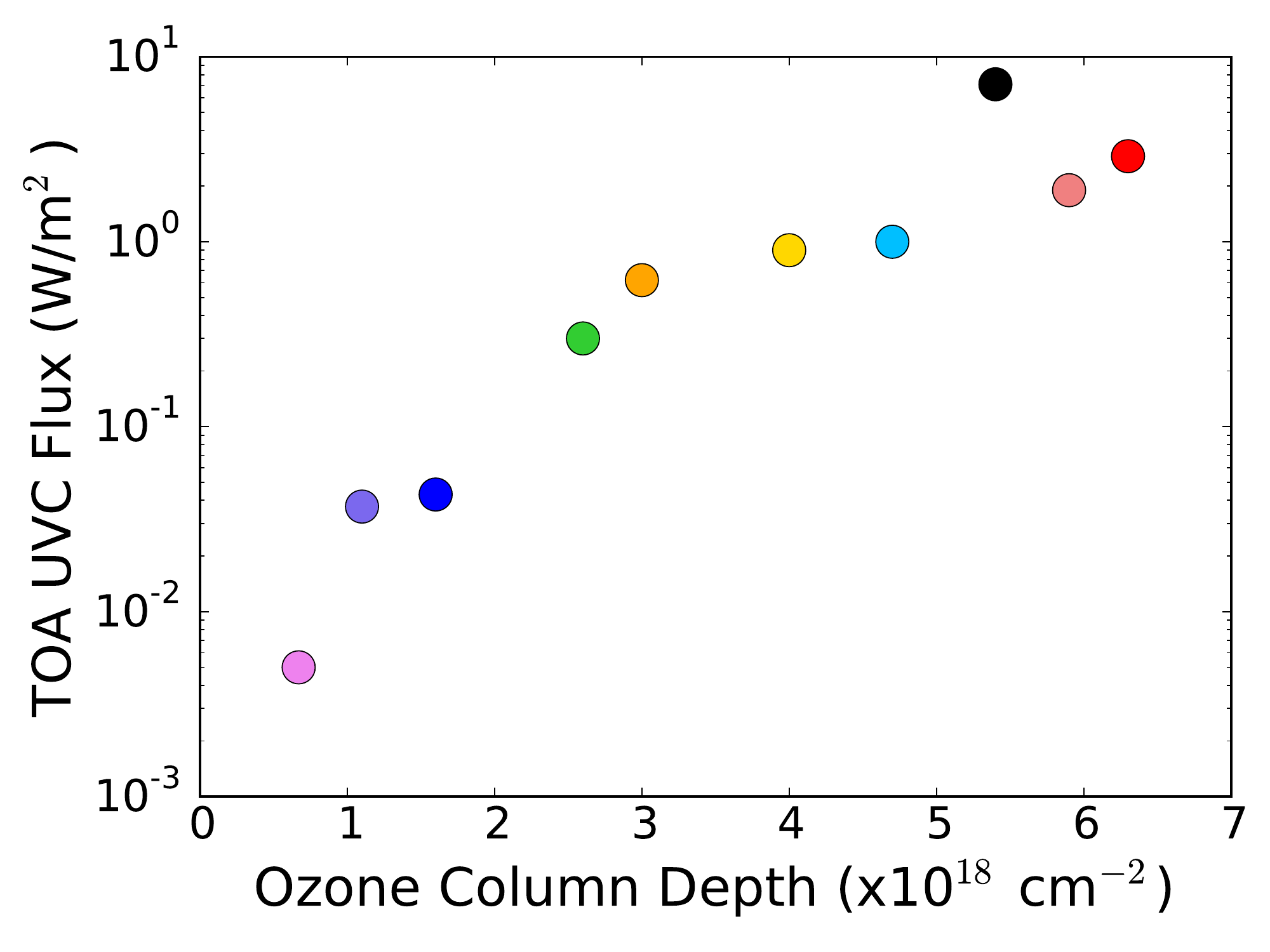} \\
\includegraphics[scale=0.4]{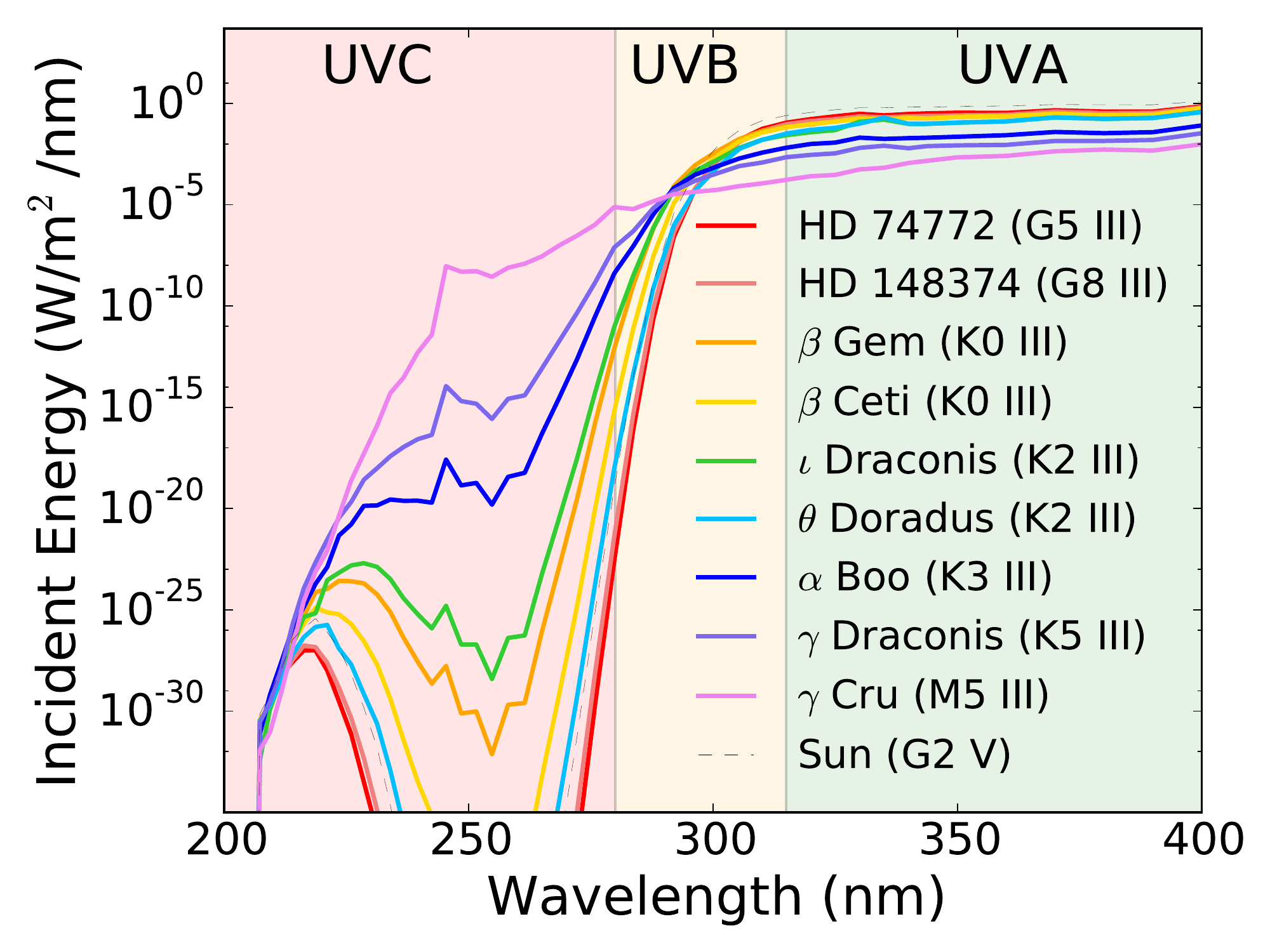}
\includegraphics[scale=0.4]{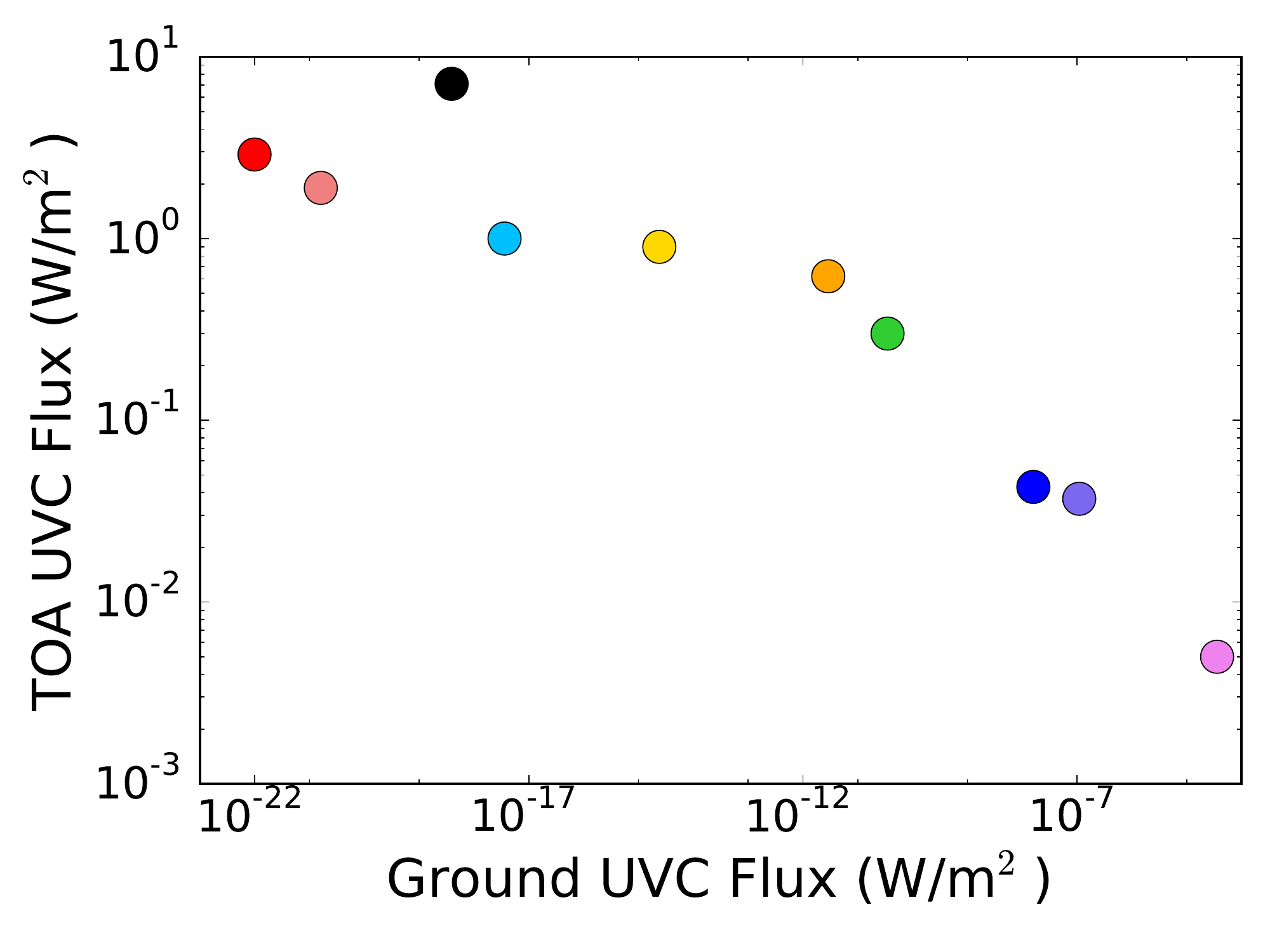}
\caption{(Top, left) Incident top of the atmosphere (TOA) integrated UVC flux versus post-MS grid stars effective temperature, (bottom, left) ground UV flux shown as a function of wavelength for Earth-like planet models, (top, right) TOA integrated UVC flux versus total ozone column depth, and (bottom, right) TOA  integrated UVC flux versus ground integrated UVC flux . The different colored dots/lines indicate individual post-MS grid host stars for the planet model, the black dots/line shows present-day Earth for comparison.  \label{UV_ground}}
\end{figure}

Planets orbiting cooler stars in the HZ can receive lower levels of UV flux (see Figure~\ref{RG_models}), impacting ozone production rates.  As seen in Equation~\ref{O3} ozone creation requires short-wavelength UV photons ($\lambda<$~240~nm), resulting in larger amounts of ozone for planets experiencing higher amounts of incident UVC (100~nm$~<\lambda<~$280~nm).  Since our target stars all have lower UVC radiation than the Sun, model atmospheres at the Earth-equivalent distance tend to have lower ozone production rates, resulting in smaller ozone column depths (Table~\ref{model_summary} and Figure~\ref{UV_ground}).  The calculated ozone column depth of the planet orbiting in the HZ of the coolest grid star is  only 11\% of the hottest grid host star planet in the HZ,and 12\% of the column depth on present-day Earth. 

This lower amount of incident UV flux causes lower photolysis rates in the upper atmosphere, with planets orbiting cooler stars experiencing less photolysis-related depletion (see Figure~\ref{profiles}).  This effect is particularly relevant for CH$_4$, N$_2$O, and CH$_3$Cl, which are also heavily depleted via reactions with OH, a by-product of ozone photolysis.  The comparable/higher levels of these species should maintain the detectability of both gases in combination as a sign of biological activity on a planet.  Our post-MS model planet photochemistry profiles are consistent with studies of cooler MS stars (e.g.\ \cite{segu05,rugh13,rugh15}).

\begin{table}[h!]
\begin{center}
\small
\caption{Model Summary for Earth-Equivalent Distances \label{model_summary}}
\begin{tabular}{llccrrr}
Spectral & Star  &Stellar $T_{\footnotesize \mbox{eff}}$ &  Surface $T_{\footnotesize \mbox{eff}}$ & UVB TOA &UVC TOA & Ozone Column   \\
Type	& Name & (K)	& (K)	& W/m$^2$	& W/m$^2$	& Depth (cm$^{-2}$)  \\
\hline
G2 V 	& Present-day Earth & 5775 & 288.2 & 18.9	&7.1	& 5.4$\times10^{18}$\\
\hline
G5 III	& HD 74772		&	5118		&	289.3 & 8.3	& 2.9		& 6.3$\times10^{18}$\\
G8 III	& HD 148374		&	4948		& 	294.5 & 5.8	& 1.9		& 5.9$\times10^{18}$\\
K0 III		& $\beta$ Gem		&	4865		&	295.0 & 3.1	& 0.62	& 3.0$\times10^{18}$\\
K0 III		& $\beta$ Ceti		&	4797		&	295.5 & 3.2	& 0.90	& 4.0$\times10^{18}$\\
K2 III		& $\iota$ Draconis	&	4445		&	298.6 & 0.96	& 0.30	& 2.6$\times10^{18}$\\
K2 III		& $\theta$ Doradus	&	4320		&	299.1 & 1.4	& 1.0		& 4.7$\times10^{18}$\\
K3 III		& $\alpha$ Boo		&	4286		&	303.6 & 0.19	& 0.043	& 1.6$\times10^{18}$\\
K5 III		& $\gamma$ Draconis &	3989		&	304.7 & 0.063	& 0.037	& 1.1$\times10^{18}$\\
M5 III	& $\gamma$ Crucis	&	3626		&	306.4 & 0.0054	& 0.0050	& 6.7$\times10^{17}$\\
\hline
\label{planets}
\end{tabular}
\end{center}
\end{table}

Table~\ref{model_summary} shows the TOA integrated flux for UVB and UVC at the 1~AU-equivalent distance, as well as the resulting ozone column depth and average surface temperature for Earth-like planetary models with 1~bar surface pressure. It also shows present-day Earth values for comparison in the first line.

The UV ground fluxes vary drastically as a function of wavelength due to the wavelength-dependent nature of ozone-shielding effectiveness.  Because we use observed UV data for our grid stars, and these stars are at different stages in their post-MS evolution, the correlation between effective temperature and UVC flux is not exact (see all UV fluxes in Table~\ref{UV_all}). However, the correlation between ozone column depth and UVC surface irradiation is clearly shown in Figure~\ref{UV_ground}. 

Although planets with cooler post-MS host stars generally have lower levels of incident UVC flux, the lower amounts also decrease ozone production, allowing a higher percentage of UVB and UVC radiation to reach their surfaces compared to planets orbiting post-MS stars with higher incident UVC flux. The differences in the attenuation percentage are most significant for UVC flux, which is almost entirely shielded by ozone. As seen in Figure~\ref{UV_ground} higher incident UVC flux correlates with higher ozone column depth and lower ground UVC as a result of ozone shielding. Note that both $\beta$ Ceti and $\theta$ Doradus have high UVC flux compared to the grid stars with similar surface temperature, and H-R diagram fitting matches both stars to a later stage in the post-MS evolution. All model planets for post-MS host stars with lower effective temperatures than the Sun show higher UVC surface levels than present-day Earth, with model planets orbiting cooler post-MS hosts generally experiencing higher levels of surface UVC than similar planets orbiting hotter post-MS hosts (see Table~\ref{UV_all}).

\begin{table}[h!]
\begin{center}
\small
\caption{UV Integrated Fluxes for Earth-Equivalent Distances\label{UV_all}}
\begin{tabular}{cl|rrr|rrr}
Spectral  & Star  & \multicolumn{3}{c}{UVB 280 - 315 nm (W/m$^2$)} & \multicolumn{3}{c}{UVC 121.6 - 280 nm (W/m$^2$)}\\
\cline{3-5} \cline{6-8}
Type & Name  & ITOA & Ground &\% to ground & ITOA & Ground &\% to ground  \\
\hline
G2 V & Present day Earth  & 18.9 & 2.2 & 11 & 7.1 & 3.9E-19 & 5.4E-18 \\
\hline
G5 III	& HD 74772			& 8.3		& 0.92	& 11	& 2.9		& 9.9E-23     & 3.4E-21  \\
G8 III	& HD 148374			& 5.8		& 0.76	& 13	& 1.9		& 1.6E-21     & 8.4E-20  \\
K0 III		& Pollux				& 3.1		& 0.69	& 22 & 0.62	& 2.9E-12     & 4.7E-10  \\
K0 III		& $\beta$ Ceti			& 3.2		& 0.59	& 19 & 0.90	& 2.4E-15     & 2.7E-13  \\
K2 III		& $\iota$ Draconis		& 0.96	& 0.26	& 27 & 0.30	& 3.5E-11     & 1.2E-8  \\
K2 III		& $\theta$ Doradus		 & 1.4   	& 0.27       & 19 & 1.0	& 3.6E-18		& 3.4E-16 \\
K3 III		& $\alpha$ Boo			 & 0.19	& 0.065     & 34 & 0.043	& 1.6E-8		& 3.6E-5  \\
K5 III		& $\gamma$ Draconis 	 & 0.063	& 0.023     & 37 & 0.037	& 3.1E-7     & 0.00084  \\
M5 III	& $\gamma$ Crucis		 & 0.0054	& 0.0025	 & 46 & 0.0050	& 3.6E-5     & 0.71  \\

\hline
\end{tabular}
\end{center}
\end{table}

\subsection{Planetary Atmospheres: post-MS evolution \label{evol_sec}}

\begin{figure}[h!]
\centering
\includegraphics[scale=0.8]{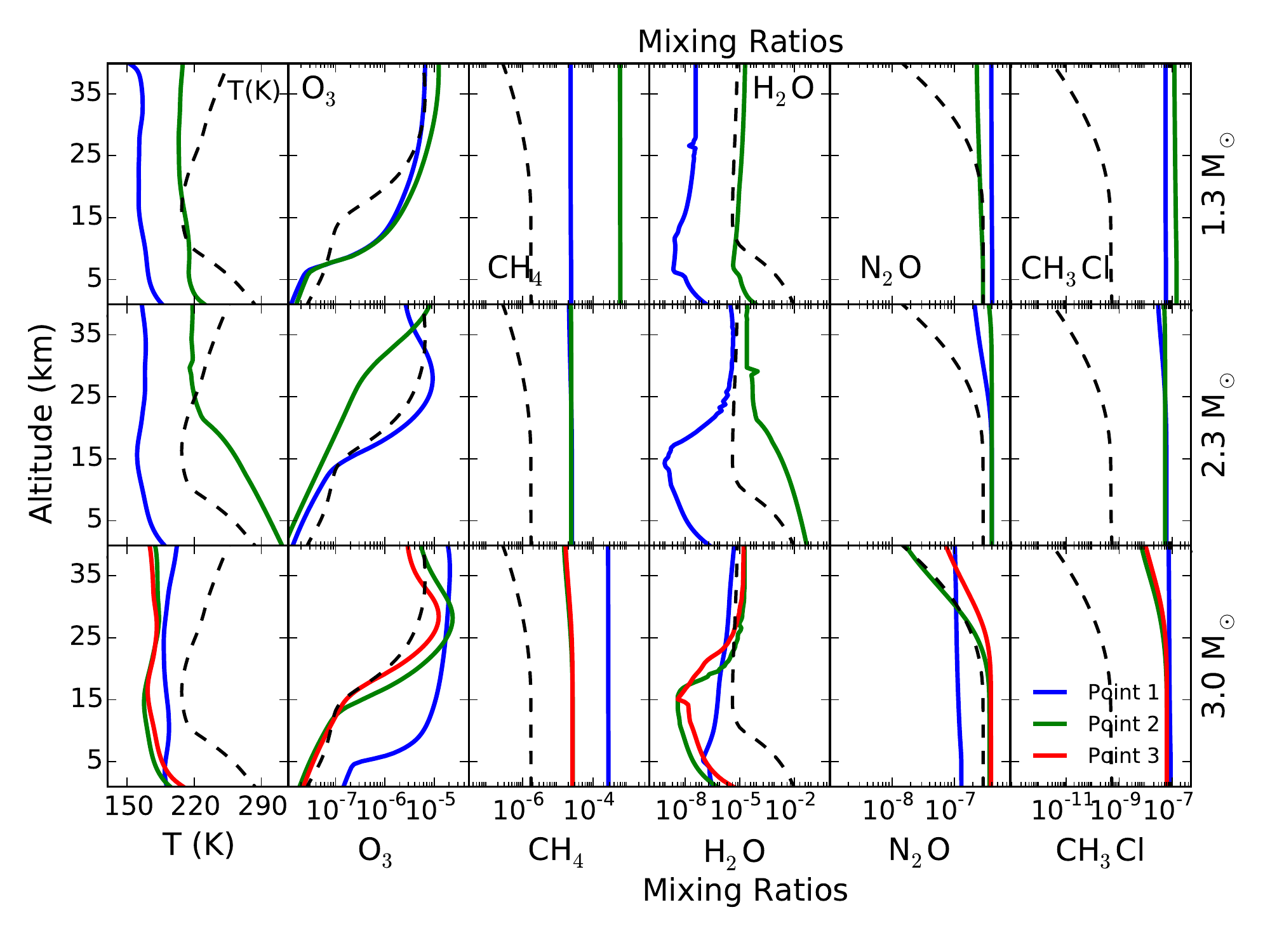}
\caption{Evolving temperature and photochemical profiles of Earth-like planets orbiting post-MS host stars on 1.3 (top), 2.3 (middle), and 3.0 (bottom) mass tracks.  Blue lines represent the first point in evolution sampled, green lines the second, and for the 3.0~\Ms\ track red lines the third.  \label{evol_profiles}}
\end{figure}

To explore a planet's characteristics while orbiting an evolving post-MS star, we model the same planet  at different points along its host star's stellar evolution. We include the increase of the initial semimajor axis of the planet, which evolves outward with stellar mass loss (see Section~\ref{HZ_dist}). We model planets orbiting post-MS host stars on three different mass tracks, with which we can match our observed post-MS grid stars to specific stellar masses and evolutionary phases (see Table~\ref{RG_targets}). The temperature and photochemical profiles of our model planets are shown in Figure~\ref{evol_profiles}, a summary of the results is shown is in Table~\ref{summary_evol}, and the UV surface flux data are summarized in Table~\ref{UV_evol} and compared to present-day Earth. 

\begin{table}[h!]
\begin{center}
\small
\caption{Model Summary throughout post-MS evolution \label{summary_evol}}
\begin{tabular}{lccrrrr}
\hline
Star & Track & Evolutionary & Semimajor & $S_{\footnotesize \mbox{eff}}$ & Surface & Ozone Column  \\
Name & Mass (\Ms) & Stage & Axis (AU) & & Pressure (bar) & Depth cm$^{-2}$  \\
\hline
$\iota$ Draconis	&	1.3 &		HB		&	13.8		&	0.2689 &	0.95 &	5.1E+18	\\
$\theta$ Doradus	&	1.3 &		HB/AGB	&	13.8		&	0.5507 &	0.95 &	1.19E+19	\\	
\hline
$\beta$ Gem		&	2.3 &		HB		&	12.0		&	0.2835 &	1.0	&	3.2E+18	\\
$\gamma$ Draconis 	&	2.3 &		HB/AGB &	12.1	&	1.0647 &	0.99 &	2.5E+18 \\
\hline
HD 74772			&	3.0 &		HB I	&	18.2	&	0.2824 &	1.00	&	5.3E+19	\\
HD 148374		&	3.0 &		HB II	&	18.2	&	0.3239 &	1.00 &	7.4E+18  \\
$\beta$ Ceti	&	3.0 &		AGB	&	18.23	&	0.4095 &	1.00 &	4.1E+18  \\
\hline
\end{tabular}
\end{center}
\end{table}

Planets that would spend the maximum overall time in the post-MS HZ orbit initially near the outer edge of the post-MS HZ, warming with increasing incident stellar flux as the luminosity of the post-MS host increases.  Note that at the outer edge of the post-MS HZ, planets initially experience similar to lower levels of UVC flux than present-day Earth, producing slightly lower amounts of ozone. Although less ozone shielding exists, the lower incident flux produced UV surface environments comparable to present-day Earth for most of our models (Figure~\ref{evol_profiles} and Table~\ref{summary_evol}). These model atmospheres produce similar to slightly lower amounts of ozone but higher amounts of methane than on present-day Earth. The N$_2$O levels are similar, while CH$_3$Cl levels are similar to slightly higher for these post-MS planetary atmospheres.  For the 1.3~\Ms\ case there is a slight decrease in UV to the ground from the first to second point in evolution sampled due to an increase in incident UV causing higher ozone production rates.  There is an increase of UV to the ground over time for the~2.3 and 3.0~\Ms\ cases resulting from lower amounts of incident UV causing less ozone shielding (see Table~\ref{UV_evol}).

\begin{figure}[h!]
\includegraphics[scale=0.4]{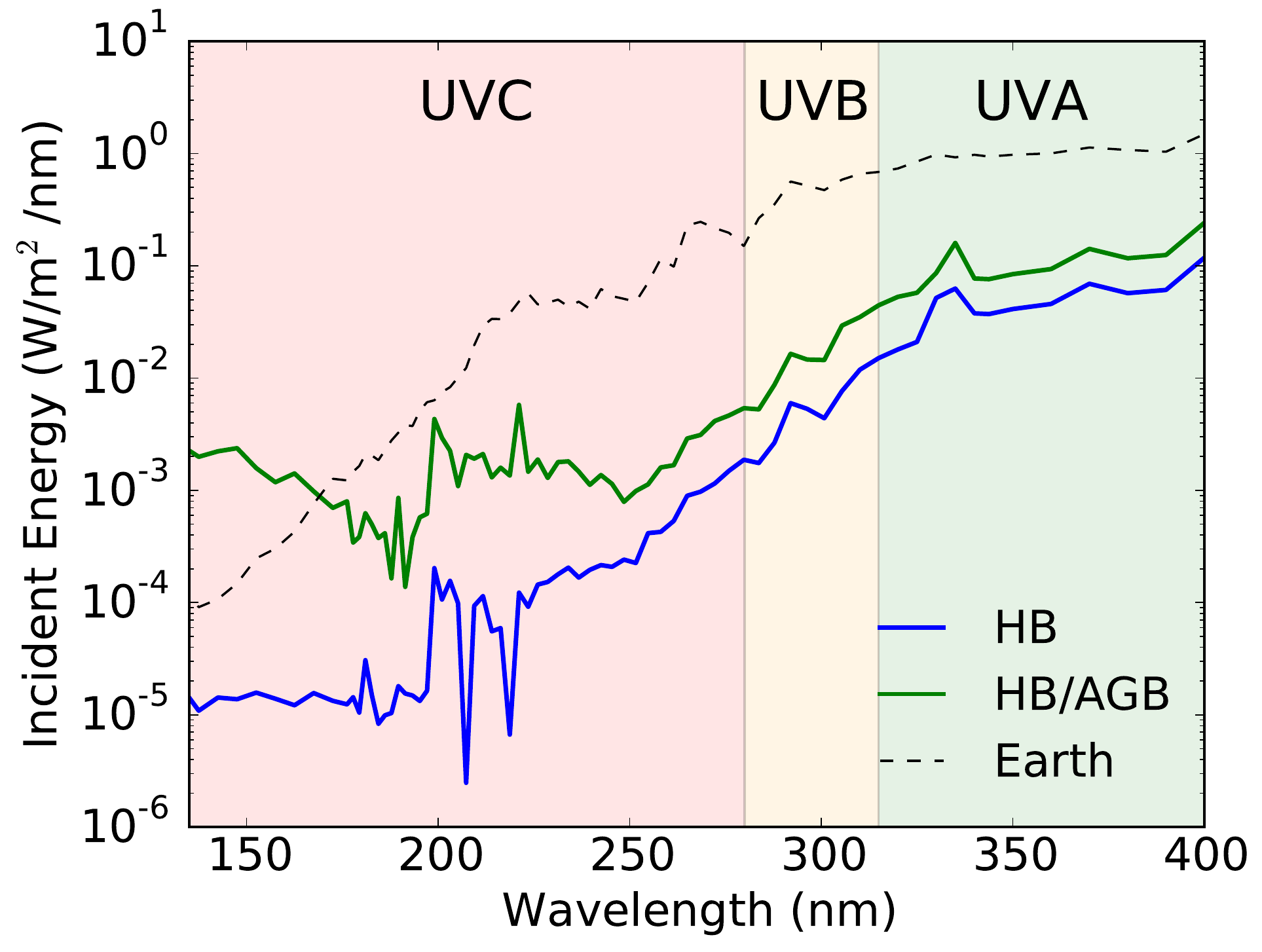}
\includegraphics[scale=0.4]{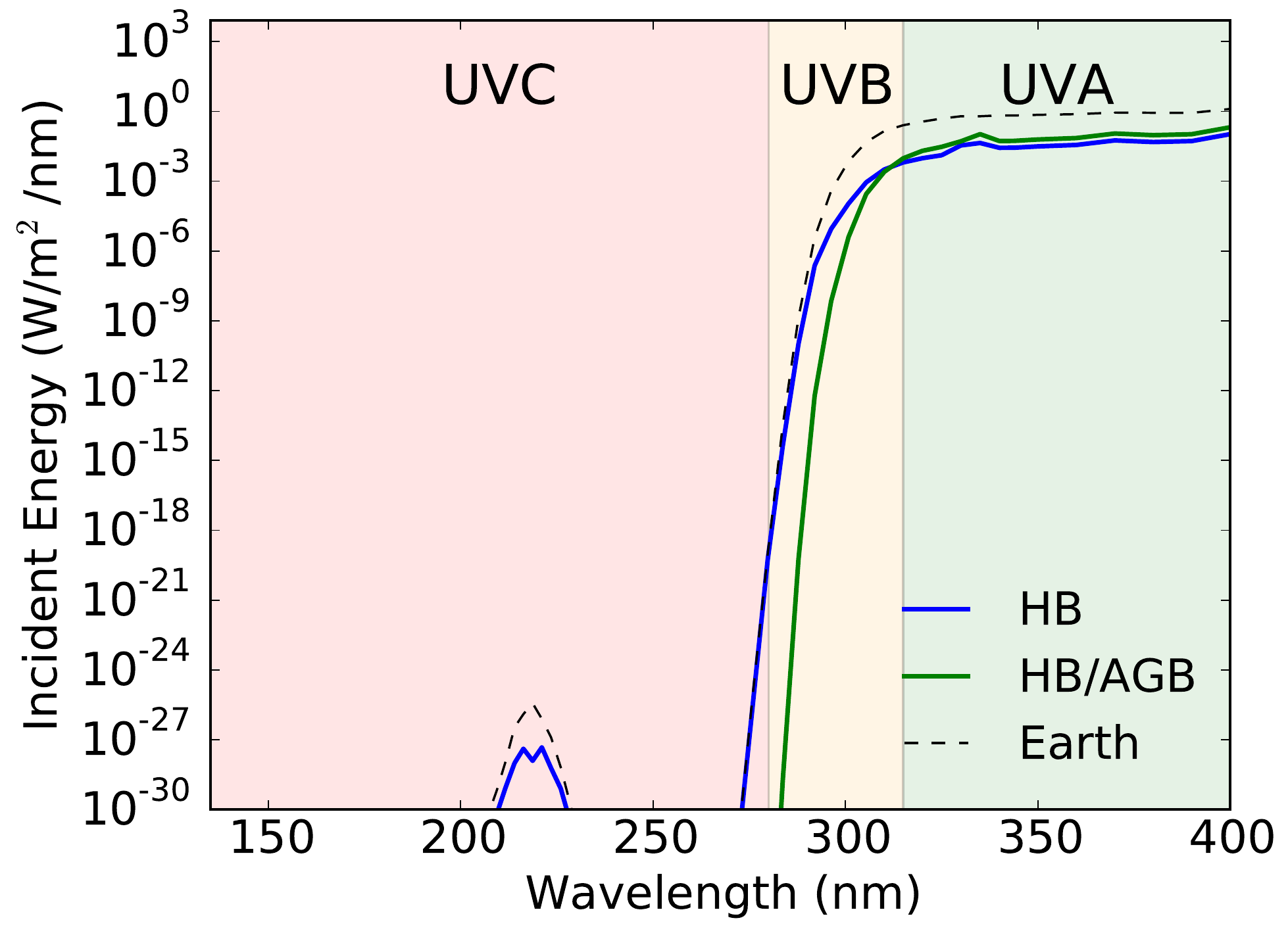}  \\
\includegraphics[scale=0.4]{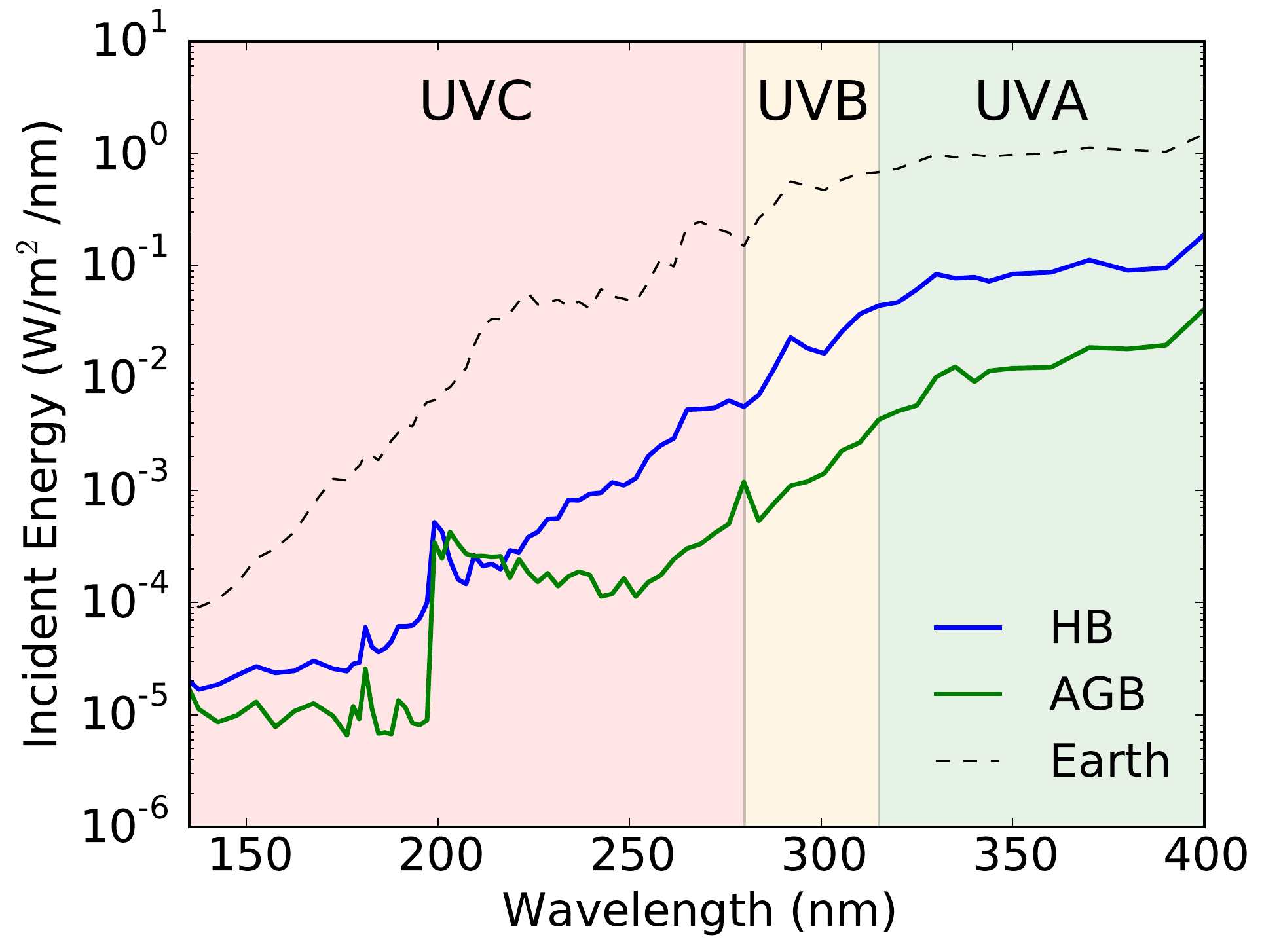}
\includegraphics[scale=0.4]{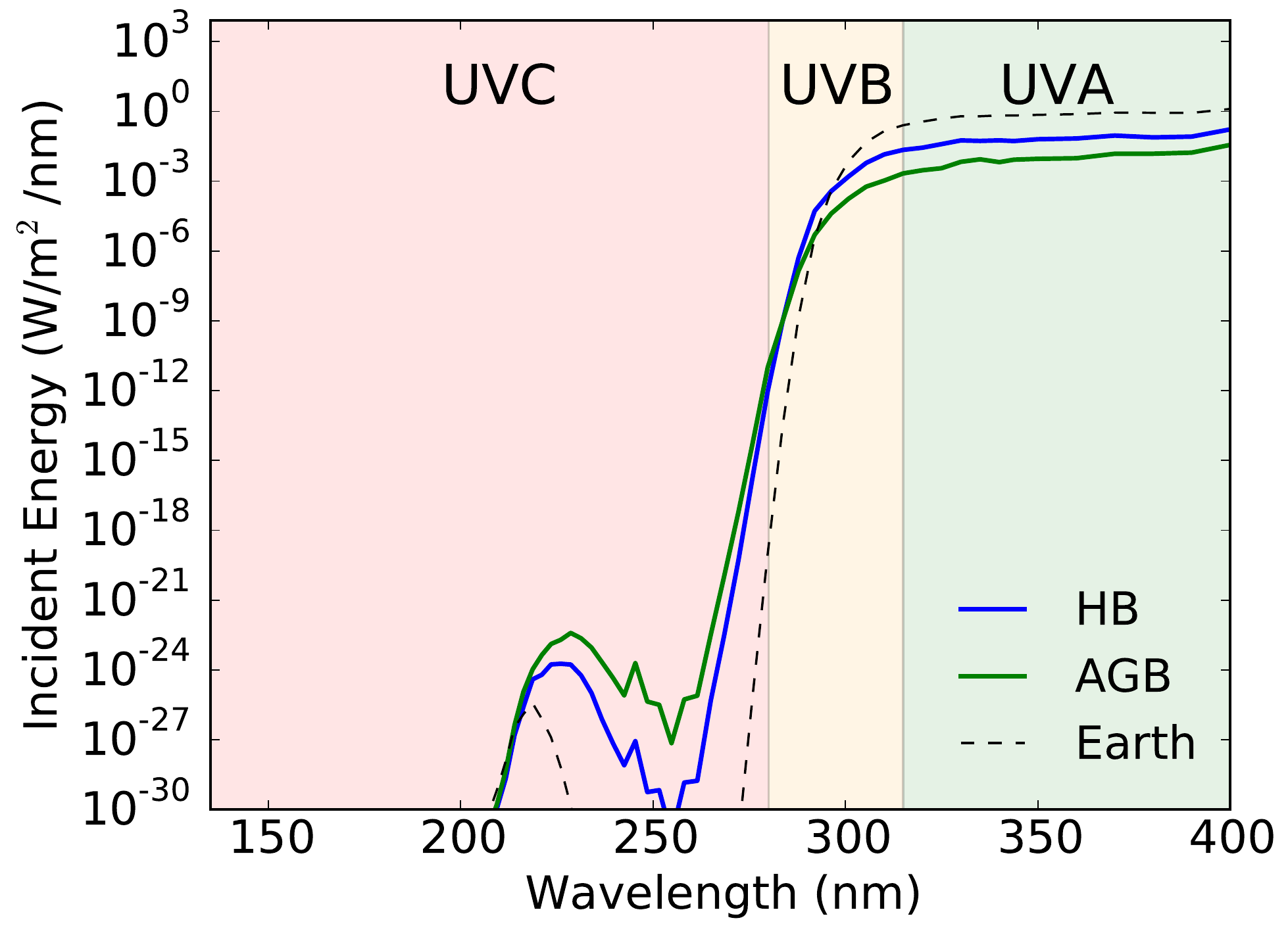}  \\
\includegraphics[scale=0.4]{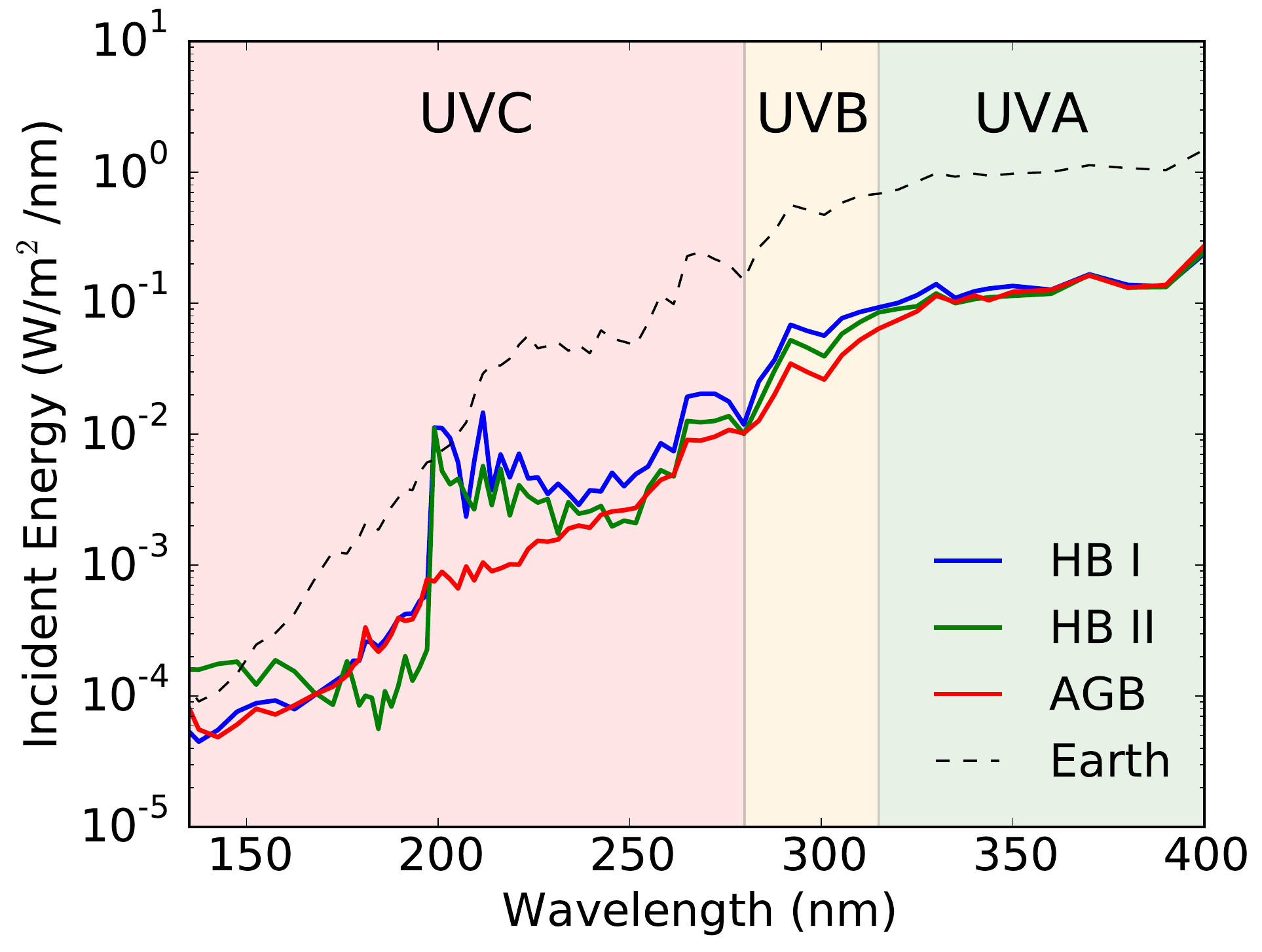}
\includegraphics[scale=0.4]{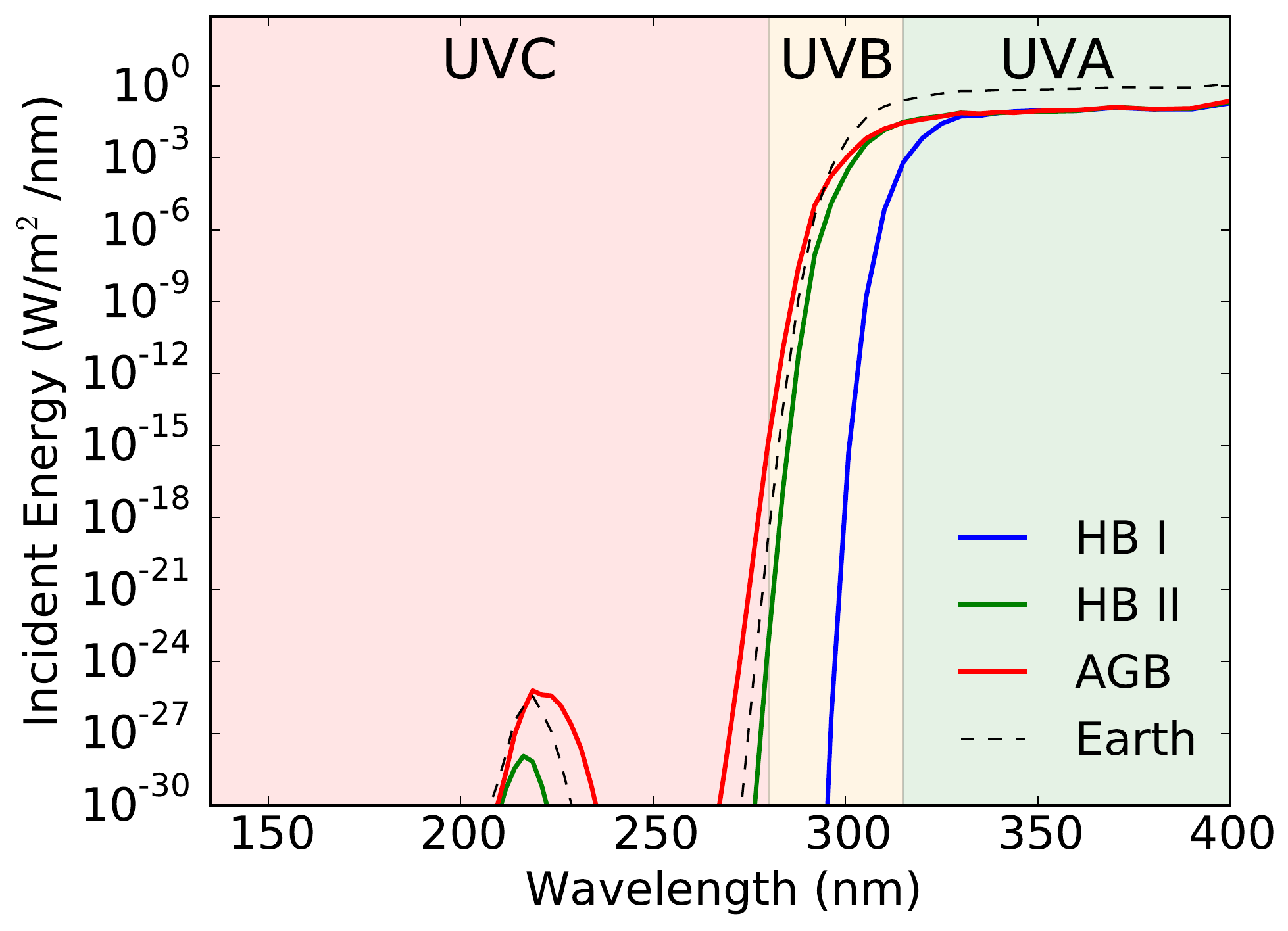}
\caption{Top of the atmosphere and surface UV fluxes for Earth-like planets in the post-MS HZ orbiting post-MS stars. We mimic their host stars evolution by using known stars matched to different parts of their post-MS evolution as marked in Figure~\ref{HZ_tracks}. \label{UV_ground_evol}}
\end{figure}

\begin{table}[h!]
\begin{center}
\caption{UV Integrated Fluxes throughout Post-MS Evolution \label{UV_evol}}
\begin{tabular}{ll|rrr|rrr}
Star & Mass & \multicolumn{3}{c}{UVB 280 - 315 nm (W/m$^2$)} & \multicolumn{3}{c}{UVC 121.6 - 280 nm (W/m$^2$)}\\
\cline{3-5} \cline{6-8}
Name& (\Ms)  & ITOA & Ground &\% to ground & ITOA & Ground &\% to ground  \\
\hline
 Present day Earth & 1.0 & 18.9 & 2.2 & 11 & 7.1 & 3.9E-19 & 5.4E-18 \\
\hline
$\iota$ Draconis	& 1.3		& 0.26	& 0.052	& 20		& 0.081	& 2.1E-19     & 2.6E-16  \\
$\theta$ Doradus	& 1.3		& 0.79	 & 0.062	& 7.8		 & 0.58	 & 3.6E-31     & 6.3E-29  \\ 
\hline
$\beta$ Gem		& 2.3		& 0.87	& 0.22        & 25	& 0.18	 & 3.7E-12     & 2.1E-9  \\ 
$\gamma$ Draconis	& 2.3		& 0.067	& 0.020	& 30		& 0.039	& 5.0E-11     & 1.3E-7 \\
\hline
HD 74772			& 3.0		& 2.3		& 0.0033	& 0.14	& 0.82	& 2.5E-46     & 3.0E-44  \\
HD 148374		& 3.0		& 1.9    	& 0.25       & 13 	& 0.60 	& 1.3E-23     & 2.1E-21 \\
$\beta$ Ceti		& 3.0		& 1.3    	& 0.27       & 21 	& 0.37 	 & 7.7E-15     & 2.1E-12\\

\hline
\end{tabular}
\end{center}
\end{table}

Unlike the profiles shown in Figure~\ref{profiles} for models at the Earth-equivalent distance, these models are mainly near the outer edge of the post-MS HZ and have lower incident UV flux and therefore lower amounts of upper atmosphere photolysis (Figure~\ref{evol_profiles}). This is most evident in the nearly vertical profiles of the species most sensitive to photolysis and its products: CH$_4$, N$_2$O, and CH$_3$Cl.

\section{Discussion \label{discussion}}

\subsection{How could life form/become detectable on objects habitable during the post-MS?}

It is uncertain how long life needs to emerge and evolve, with estimates ranging from~0.5 to 1~Gyr (see, e.g.\ \cite{furn04,lope05}).  However, even if post-MS HZ timescales would be too short for the development of life, increasing luminosity during the RGB phase should melt the initially frozen surfaces of moons and planets, revealing subsurface life \citep{rami16}. Frozen planets and moons are not excluded from hosting life; however, it is unknown whether subsurface biospheres--for example, under an ice layer on a frozen planet--can modify a planet's atmosphere in ways that can be detected remotely. Thus, we concentrate on the liquid water zone in our paper.

\subsection{Could we directly image planets in the post-MS HZ of RGs?}

Upcoming ground-based ELTs are also designed to characterize exoplanets. Even though a post-MS host star is very bright compared to an MS star, the wider separation of the post-MS HZ when compared to an MS star's HZ also opens up the possibility of directly imaging planets in the post-MS HZ for close-by systems (see \cite{lope05,rami16}).  To further explore this idea, we have compiled a list of post-MS luminosity class III stars within 30~pc and indicate whether these stars have available IUE data (see Table~\ref{nearby}). We compiled this target list using several published lists \citep{lope05,luck15,stoc18} and updated their distances using \emph{Gaia} DR2 data \citep{gaia16}.

\begin{table}[h!]
\scriptsize
\caption{Luminosity class III stars within 30 pc \label{nearby}}
\begin{center}
\begin{tabular}{lllcrrrrrrrr}
\hline
Mass & Spectral & Star & IUE & Stellar & Distance & Radius & 1 AU equiv.\  & Ang. Sep.\ & Post-MS & HB  \\
(\Ms) & Type & Name & data$^1$ & $T_{\footnotesize \mbox{eff}}$ (K) & (pc) & (R$_\odot$) &   (AU) & (") & (Myr) &(Myr) \\
 \hline
 2.0$^*$	&	K1.5 III	& $\alpha$ Bo\"otis	& B	& 4286	& 11.3	& 25.4	& 13.98	& 1.24	&	260	& 206	\\
 2.0$^*$	&	M3.5 III	& $\gamma$ Crucis	& B	& 3626	& 27.2	& 84		& 33.08	& 1.22	&	260	& 206	\\
 1.3	&	K5 III		& $\alpha$ Tauri		& N	& 3910	& 20.5	& 44.13	& 20.21	& 0.99	&	633	& 124	\\
2.3$^*$	&	K0 III		& $\beta$ Gem 		& B	& 4865	& 10.4	& 8.8		& 6.24	& 0.60	& 	260	& 226  	\\
1.7	&	K2 III		& $\alpha$ Arietis		& S	& 4480	& 20.2	& 14.9	& 8.96	& 0.44	&	236	& 136	\\
2.5	&	K0 III		& $\theta$ Centauri 		& S	& 4980	& 18.0	& 10.6	& 7.87	& 0.44	&	201	& 185	\\
3.0$^*$	&	K0 III		& $\beta$ Ceti		& B	& 4797	& 29.5	& 16.78	& 11.57	& 0.39	&	104	& 92		\\
1.5	&	K2.5 III	& $\epsilon$ Scorpii		& S	& 4560	& 23.2	& 12.6	& 7.85	& 0.34	&	294	& 125	\\
2.3	&	K0 III		& $\gamma^2$	Sagittarii	& S	& 4769	& 24.9	& 12		& 8.17	& 0.33	&	260	& 226 	\\
1.9	&	K0 III		& $\epsilon$ Cygni		& N	& 4710	& 23.2	& 10.82	& 7.19	& 0.31	&	238	& 171	\\
3.0	&	G8 III 	& $\eta$ Draconis 		& S	& 5055	& 27.3	& 11		& 8.42	& 0.31	 & 	104	& 92		\\
1.3	&	K2 III		& $\beta$ Ophiuchi		& S	& 4467	& 24.4	& 12.42	& 7.42	& 0.30	&	663	& 124	\\ 
1.5	&	K2 III		& $\kappa$ Ophiuchi		& S	& 4529	& 27.2	& 11		& 6.76	& 0.25	&	294	& 125 	\\
1.5	&	K2 III		& $\beta$ Columbae		& N	& 4545	& 27.8	& 11.5	& 7.12	& 0.26	&	294	& 125 	\\
1.8	&	A7 III		& $\gamma$ Bo\"otis	& S	& 7800	& 26.7	& 3.34	& 6.09	& 0.23	&	157	& 155	\\
1.7	&	K0 III		& $\nu$ Octantis		& S	& 4860	& 19.4	& 5.81	& 4.11	& 0.21	&	236	& 136	\\
1.3	&	K1 III		& $\nu^2$ CMa			& S	& 4790	& 19.8	& 4.9		& 3.37	& 0.17	&	663	& 124 	\\
1.8	&	 A5 III	& 111 Her				& N	& 8153	& 28.3	& 1.83	& 3.64	& 0.13	& 	157	& 155	\\
 \hline
  \end{tabular}  \\
  $^1$Key for the IUE data: N = no data, S = short wave data, B = both short wave and long wave data.  \\
  *Indicates that target is modeled in this paper
 \end{center}
 \end{table}

Comparing the angular separation ($\theta$ (arcsec)~=~$a$~(AU) / $d$~(pc); where $a$~=~planet semi-major axis, $d$~=~distance to star system) for the angular separation of the 1~AU equivalent distance (Table~\ref{nearby}) with the inner working angle (IWA) for a telescope, which describes the minimum angular separation at which a faint object can be detected around a bright star, we can determine whether planets at such orbital distances could be remotely detected and resolved in the near future. For example, the 38~m diameter Extremely Large Telescope should have an IWA of about 6~milliarcseconds (mas) when observing in the visible region of the spectrum (assuming $\theta$~IWA $\approx 2(\lambda D$), where $\lambda$ is the observing wavelength and $D$ is the telescope diameter). Table~\ref{nearby} shows the angular separation of their 1 AU equivalent orbital distance to compare them to predicted IWAs for ground-based telescopes like the ELT with a proposed IWA of 6~milliarcseconds \citep{quan15}.  Note that we excluded binary systems and variables from our list.  The angular separation of the 1~AU equivalent orbital distance during the post-MS for nearby RGs ranges from~1.2 to 0.2~arcseconds, making such observations interesting for upcoming ground-based direct imaging surveys.

\subsection{Continuous time in the HZ}

Note that Table~\ref{HZ_time} shows two values for the maximum time a planet orbiting a post-MS host star can stay in the HZ: the time a planet can stay continuously in the post-MS CHZ and a longer time period that consists of the sum of the time a planet can spend in the HZ during a star's post-MS evolution. The sum of time a planet can spend in the post-MS HZ is interrupted by the times a planet receives higher luminosity, which should put it into a runaway greenhouse state (see, e.g.\ \cite{kast93}), where a planet would boil off surface water and subsequently would lose hydrogen to space. Estimates of the time such water loss would take vary (see, e.g.\  \cite{kast93,kopp13}). Whether such a runaway greenhouse would be reversible if the irradiation of the star decreases during the HB phase is unclear and will depend on many factors, including cloud feedback during the onset of the runaway greenhouse stage, which is unknown and will only be constrained once atmospheric observations of such planets are available. Thus, Table~\ref{HZ_time} shows both values.

\subsection{Comparisons to previous studies}

Earlier studies have explored the post-MS HZ boundaries and evolution (e.g.\ \cite{rami16,danc13,ster03}), however, they did not explore planetary climate and UV surface environments on planets in the post-MS HZ. Here we compare our results of post-MS HZ boundary evolution to earlier studies. Note that \cite{rami17,rami18} have expanded the HZ concept to stars with higher surface temperatures, which allows an extension of post-MS HZ calculations to higher-mass stars. We calculate post-MS HZ limits, stellar mass-loss rates, and planetary semimajor axis evolution following \cite{rami16}, and find consistent results for the overlapping mass range between the two studies (1.0 to 1.9~\Ms). \cite{ster03} calculates temperate distances and habitability timescales for 1 to 3~\Ms\ post-MS stars, both of which are consistent with our HZ limits. We also find multiple periods of habitability for a range of semimajor axes during the post-MS, as noted by \cite{danc13}.

\section{Conclusions \label{conclusions}}

We explore how long planets could remain in the post-MS HZ for post-MS stars from~1.0 to 3.5~\Ms. We additionally study how the stellar environment would affect the atmospheric composition and potentially detectable biosignatures, as well as the surface UV conditions of Earth-mass planets.  We model atmospheric erosion and semimajor axis evolution resulting from stellar winds/mass loss.

Less massive grid stars that do undergo a helium flash ($<$~2~\Ms) experience a larger luminosity increase during the post-MS (Table~\ref{HZ_limits}) causing a more drastic change in the orbital distance of the post-MS HZ compared to the MS HZ, as well as higher stellar mass-loss rates and semimajor axis variation.  We find that the maximum time a planet can spend in the post-MS HZ is between~56 and 257~Myr, for our grid stars, which is highly dependent on the amount of time the host star spends on the relatively stable HB. The wide orbital separation of the post-MS HZ limits atmospheric erosion in our model to about 10\% of the initial atmosphere for Earth-like planets with 1~bar surface pressure.   Our models are consistent with mass cases from \cite{rami16} that also use the Padova catalog ($\ge$~1~\Ms).

Model planet atmospheres at the Earth-equivalent distance orbiting evolved stars (see Figures~\ref{profiles} and~\ref{UV_ground}) produce lower amounts of ozone but higher amounts of methane than on present-day Earth, which should maintain the detectability of both gases in combination as a sign of biological activity on a planet. The N$_2$O levels are similar, while the CH$_3$Cl levels are similar to slightly higher in the model atmosphere of Earth-like planets orbiting post-MS stars receiving similar irradiation than Earth does from our Sun.

Lower ozone production results in higher amounts of UVC reaching the surface of our model planets (see Figures~\ref{profiles} and~\ref{UV_ground}). However, planets on orbits that maximize the overall time in the post-MS HZ spend extended periods of time close to the outer edge of the post-MS HZ, with less incident UVC flux and UVC surface levels comparable to present-day Earth (Figures~\ref{evol_profiles} and~\ref{UV_ground_evol}).

Although the HZ timescales for post-MS stars may not be sufficient for life to develop and evolve, increased luminosity levels cause the post-MS HZ to move past the system's original frost line, potentially melting previously icy planets or moons to reveal subsurface life. In addition, the wider separation of post-MS HZs results in a larger host-planet apparent angular separation, making them interesting targets for direct imaging with upcoming large telescopes, extending the search for habitable planets to older planetary systems.

\acknowledgments
We thank our anonymous referee for their comments that made our manuscript clearer.  L.K.\ and T.K.\ acknowledge support from the Simons Foundation (SCOL \# 290357, Kaltenegger) and the Carl Sagan Institute.  T.K. was additionally funded by a NASA Space Grant Fellowship.


\clearpage
\begin{thebibliography}{}

\bibitem[Baud \& Habing(1983)]{baud83} Baud, B., \& Habing, H.~J.\ 1983, \aap, 127, 73

\bibitem[Barnes \& Heller(2013)]{barn13} Barnes, R., \& Heller, R.\ 2013, Astrobiology, 13, 279

\bibitem[Bertelli et al.(2009)]{bert09} Bertelli, G., Nasi, E., Girardi, L., \& Marigo, P.\ 2009, \aap, 508, 355

\bibitem[Bertelli et al.(2008)]{bert08} Bertelli, G., Girardi, L., Marigo, P., \& Nasi, E.\ 2008, \aap, 484, 815

\bibitem[Canto \& Raga(1991)]{cant91} Canto, J., \& Raga, A.~C.\ 1991, \apj, 372, 646

\bibitem[Chapman(1930)]{chap30} Chapman, S. A. (1930) Theory of Upper-Atmospheric Ozone, Mem.\ R.\ Met. Soc., 3, 26, pp 103-125

\bibitem[Danchi \& Lopez(2013)]{danc13} Danchi, W.~C., \& Lopez, B.\ 2013, \apj, 769, 27

\bibitem[Furnes et al.(2004)]{furn04} Furnes, H., Banerjee, N.~R., Muehlenbachs, K., Staudigel, H., \& de Wit, M.\ 2004, Science, 304, 578

\bibitem[Gaia Collaboration et al.(2016)]{gaia16} Gaia Collaboration, Prusti, T., de Bruijne, J.~H.~J., et al.\ 2016, \aap, 595, A1

\bibitem[Haqq-Misra et al.(2008)]{haqq08} Haqq-Misra, J.~D., Domagal-Goldman, S.~D., Kasting, P.~J., \& Kasting, J.~F.\ 2008, Astrobiology, 8, 1127

\bibitem[Jones et al.(2014)]{jone14} Jones, M.~I., Jenkins, J.~S., Bluhm, P., Rojo, P., \& Melo, C.~H.~F.\ 2014, \aap, 566, A113

\bibitem[Kaltenegger(2017)]{kalt17} Kaltenegger, L.\ 2017, \araa, 55, 433

\bibitem[Kaltenegger \& Haghighipour(2013)]{kalt13} Kaltenegger, L., \& Haghighipour, N.\ 2013, \apj, 777, 165

\bibitem[Kaltenegger(2010)]{kalt10} Kaltenegger, L.\ 2010, \apjl, 712, L125

\bibitem[Kane \& Hinkel(2013)]{kane13} Kane, S.~R., \& Hinkel, N.~R.\ 2013, \apj, 762, 7

\bibitem[Kasting et al.(1993)]{kast93} Kasting, J.~F., Whitmire, D.~P., \& Reynolds, R.~T.\ 1993, \icarus, 101, 108

\bibitem[Kasting \& Ackerman(1986)]{kast86} Kasting, J.~F., \& Ackerman, T.~P.\ 1986, Science, 234, 1383 

\bibitem[Kasting et al.(1985)]{kast85}Kasting, J.~F., Holland, H.~D., Pinto, J.~P. 1985, J. Geophys. Res. 90, 10497?10510

\bibitem[Kopparapu et al.(2014)]{kopp14} Kopparapu, R.~K., Ramirez, R.~M., SchottelKotte, J., et al.\ 2014, \apjl, 787, L29

\bibitem[Kopparapu et al.(2013)]{kopp13} Kopparapu, R.~K., Ramirez, R., Kasting, J.~F., et al.\ 2013, \apj, 765, 131

\bibitem[Kozakis et al.(2018)]{koza18} Kozakis, T., Kaltenegger, L., \& Hoard, D.~W.\ 2018, \apj, 862, 69

\bibitem[Leconte et al.(2013)]{leco13} Leconte, J., Forget, F., Charnay, B., et al.\ 2013, \aap, 554, A69

\bibitem[Lopez et al.(2005)]{lope05} Lopez, B., Schneider, J., \& Danchi, W.~C.\ 2005, \apj, 627, 974

\bibitem[Luck(2015)]{luck15} Luck, R.~E.\ 2015, \aj, 150, 88

\bibitem[O'Malley-James \& Kaltenegger(2017)]{omal17} O'Malley-James, J.~T., \& Kaltenegger, L.\ 2017, \mnras, 469, L26

\bibitem[Pavlov \& Kasting(2002)]{pavl02} Pavlov, A.~A., \& Kasting, J.~F.\ 2002, Astrobiology, 2, 27

\bibitem[Pavlov et al.(2000)]{pavl00} Pavlov, A.~A., Kasting, J.~F., Brown, L.~L., Rages, K.~A., \& Freedman, R.\ 2000, \jgr, 105, 11981

\bibitem[Pickles(1998)]{pick98} Pickles, A.~J.\ 1998, \pasp, 110, 863

\bibitem[Quanz et al.(2015)]{quan15} Quanz, S.~P., Crossfield, I., Meyer, M.~R., Schmalzl, E., \& Held, J.\ 2015, International Journal of Astrobiology, 14, 279

\bibitem[Ramirez(2018)]{rami18r} Ramirez, R.~M.\ 2018, Geosciences, 8, 280 

\bibitem[Ramirez \& Kaltenegger(2018)]{rami18} Ramirez, R.~M., \& Kaltenegger, L.\ 2018, \apj, 858, 72

\bibitem[Ramirez \& Kaltenegger(2017)]{rami17} Ramirez, R.~M., \& Kaltenegger, L.\ 2017, \apjl, 837, L4

\bibitem[Ramirez \& Kaltenegger(2016)]{rami16} Ramirez, R.~M., \& Kaltenegger, L.\ 2016, \apj, 823, 6

\bibitem[Ramirez \& Kaltenegger(2014)]{rami14} Ramirez, R.~M., \& Kaltenegger, L.\ 2014, \apjl, 797, L25

\bibitem[Reimers(1975)]{reim75} Reimers, D.\ 1975, Problems in stellar atmospheres and envelopes., 229

\bibitem[Rugheimer \& Kaltenegger(2018)]{rugh18} Rugheimer, S., \& Kaltenegger, L.\ 2018, \apj, 854, 19

\bibitem[Rugheimer et al.(2015)]{rugh15} Rugheimer, S., Kaltenegger, L., Segura, A., Linsky, J., \& Mohanty, S.\ 2015, \apj, 809, 57

\bibitem[Rugheimer et al.(2015b)]{rugh15b} Rugheimer, S., Segura, A., Kaltenegger, L., \& Sasselov, D.\ 2015, \apj, 806, 137

\bibitem[Rugheimer et al.(2013)]{rugh13} Rugheimer, S., Kaltenegger, L., Zsom, A., Segura, A., \& Sasselov, D.\ 2013, Astrobiology, 13, 251

\bibitem[Segura et al.(2007)]{segu07} Segura, A., Meadows, V.~S., Kasting, J.~F., Crisp, D., \& Cohen, M.\ 2007, \aap, 472, 665

\bibitem[Segura et al.(2005)]{segu05} Segura, A., Kasting, J.~F., Meadows, V., et al.\ 2005, Astrobiology, 5, 706

\bibitem[Segura et al.(2003)]{segu03} Segura, A., Krelove, K., Kasting, J.~F., et al.\ 2003, Astrobiology, 3, 689

\bibitem[Stern(2003)]{ster03} Stern, S.~A.\ 2003, Astrobiology, 3, 317

\bibitem[Stock et al.(2018)]{stoc18} Stock, S., Reffert, S., \& Quirrenbach, A.\ 2018, \aap, 616, A33

\bibitem[Toon et al.(1989)]{toon89} Toon, O.~B., McKay, C.~P., Ackerman, T.~P., \& Santhanam, K.\ 1989, \jgr, 94, 16287

\bibitem[Turbet et al.(2017)]{turb17} Turbet, M., Forget, F., Head, J.~W., \& Wordsworth, R.\ 2017, \icarus, 288, 10

\bibitem[Vassiliadis \& Wood(1993)]{vass93} Vassiliadis, E., \& Wood, P.~R.\ 1993, \apj, 413, 641

\bibitem[Wolf(2018)]{wolf18} Wolf, E.~T.\ 2018, \apjl, 855, L14

\bibitem[Wolszczan \& Frail(1992)]{wols92} Wolszczan, A., \& Frail, D.~A.\ 1992, \nat, 355, 145

\end{thebibliography}
\end{document}